\documentclass[11pt]{article}
\usepackage[utf8]{inputenc}
\usepackage{amsfonts,epsfig}
\usepackage{stackengine}
\usepackage{subcaption}
\RequirePackage{amsthm,amsmath,amsfonts,amssymb, MnSymbol}
\usepackage{mathtools}
\usepackage{xcolor}
\graphicspath{{fig/}}
\usepackage{relsize}
\usepackage{fancyvrb}
\usepackage{geometry}
\usepackage[font={it}]{caption}
\usepackage{subcaption}
\usepackage[normalem]{ulem}
\usepackage{titlesec}
\usepackage{booktabs}
\usepackage{tikz}
\usetikzlibrary{arrows.meta}
\definecolor{pisblue}{HTML}{1F4E79}
\definecolor{pisorange}{HTML}{BF5700}
\usepackage{tabularx}
\titleformat{\section}
  {\sffamily\bfseries\upshape\large\raggedright}
  {\thesection.}{0.6em}{}
\titleformat{\subsection}
  {\sffamily\bfseries\upshape\normalsize\raggedright}
  {\thesubsection.}{0.5em}{}
\titleformat{\subsubsection}
  {\sffamily\bfseries\upshape\normalsize}
  {\thesubsubsection.}{0.5em}{}
\titleformat{\paragraph}[runin]
  {\sffamily\mdseries\itshape\normalsize}
  {\theparagraph.}{0.4em}{}[\ignorespaces]

\usepackage[round]{natbib}
\usepackage[T1]{fontenc}
\usepackage[defaultsans]{lato} 
\usepackage{inconsolata}
\usepackage{hyperref}
\hypersetup{
	colorlinks=true,
	linkcolor=black,
	citecolor=black,
	filecolor=black,
	urlcolor=black,
}
\makeatletter
\def\@maketitle{%
	\begin{center}%
		\let \footnote \thanks
		{\large \sffamily\fontseries{eb}\selectfont \@title \vspace{.125in}\par}
		{\normalsize
			\begin{tabular}[t]{c}%
				\@author
			\end{tabular}\par}%
             \vspace{0.1in}%
		{\small \@date}%
         \vspace{-0.125in}%
	\end{center}%
}
\makeatother

\definecolor{darkred}{RGB}{139,0,0}
\definecolor{darkgreen}{RGB}{0,100,0}
\definecolor{pisgreen}{HTML}{009E73}
\definecolor{pisred}{HTML}{E52521}
\usetikzlibrary{arrows.meta}

\usepackage{enumitem}
\usepackage{bbm}

\usepackage{algorithm}
\usepackage{algpseudocode}
\algrenewcommand\algorithmicrequire{\textbf{Input:}}
\algrenewcommand\algorithmicensure{\textbf{Output:}}

\usepackage{enumitem}
\newtheorem{lemma}{Lemma}

\newtheorem{theorem}{Theorem}

\newtheorem{fact}{Fact}

\DeclareMathOperator{\E}{\mathbb{E}}
\DeclareMathOperator{\Var}{\mathrm{Var}}
\DeclareMathOperator{\R}{\mathbb{R}}
\DeclareMathOperator{\n}{\mathrm{normal}}

\title{Post-stratified importance sampling}
\author{ 
 Yuling Yao \footnote{Department of Statistics and Data Science, University of Texas at Austin}
\and
Sergei Tikhonov $^*$
}
\date{ \today }

\begin{document}\sloppy
	\maketitle
	\thispagestyle{empty}
	\begin{abstract}
Importance sampling is a building block of modern statistical computing, but the weighted estimate is prone to high or infinite variance. 
Moreover, its empirical standard error is often unreliable, leaving importance sampling without practical uncertainty quantification.  Motivated by multilevel regression and poststratification (MRP) in survey sampling,  we propose post-stratified importance sampling (PIS) for estimating target expectations from proposal draws.
PIS reformulates arbitrarily high-dimensional weighting into two one-dimensional probabilistic modeling problems: regression for the conditional mean given the importance ratio and marginal density estimation for the importance ratio. Different modeling choices yield a family of PIS estimators, encompassing ordinary importance sampling as a special case. We develop a model-based implementation that combines one-dimensional Gaussian process regression for the conditional mean and score-informed heavy-tailed density estimation for the marginal law, using both the observed weights and the available density evaluations. We study the oracle gains of PIS and establish consistency under general and model-specific conditions.  Simulations and applications to cross-validation demonstrate improvements in estimation accuracy and uncertainty quantification.

\vspace{.05in}
\textbf{Keywords}: 
Bayesian computation,
importance sampling,
heavy-tailed density,
poststratification,
score-based Monte Carlo,
uncertainty quantification
\end{abstract}
	
	\section{Introduction}

\paragraph{The task of importance sampling.}
Consider the general task of computing expectations: we are given a target density $p(\theta)$ on $\theta\in\R^d$, and our goal is to compute the target expectation of an arbitrary scalar function $h(\theta)$,
\begin{equation}\label{eq:task}
\mu= \E_{p} [h(\theta)] \coloneqq \int_{\R^d} h(\theta) p(\theta) d\theta.    
\end{equation}
Instead of sampling from the target $p$, we draw an independent and identically distributed (IID) sample from a given proposal density $q$,
\[
\theta_{1}, \theta_2, \ldots, \theta_S \sim q(\theta).
\]

In practice, target and proposal densities may be known only up to normalizing constants. We write $p$ and $q$ for the evaluable, possibly unnormalized density functions, and use $\E_p[\cdot]$ for expectation under the normalized distribution proportional to $p$.  Define the \emph{importance ratio} function
\[
w= r(\theta)\coloneqq \frac{p(\theta)}{q(\theta)}.
\]
Here we denote $r(\cdot):\mathbb R^d\to[0,\infty)$ as a deterministic function of $\theta$, and $w$ as the induced scalar random variable.  We also write $w_s = r(\theta_s)$, the $s$-th realized  importance weight sample.

Depending on whether $p$ and $q$ are normalized,  the classical importance sampling (IS) or self-normalized importance sampling estimate of the target expectation $\mu$ is the weighted sample mean
\begin{equation}\label{eq:IS}
\E_p [h(\theta)]
=
\begin{cases}
\displaystyle
\hat \mu_{\mathrm{IS}}\coloneqq
S^{-1}\sum_{s=1}^S w_s h(\theta_s),
& \text{when~normalized}, \\[1.2em]
\displaystyle
\hat \mu_{\mathrm{SNIS}}\coloneqq
\frac{\sum_{s=1}^S w_s h(\theta_s)}
{\sum_{s=1}^S w_s},
& \text{when~unnormalized}.  
\end{cases}
\end{equation}

\paragraph{Why importance sampling is still important, and why it is fragile.}
Beyond its use as a stand-alone estimator, importance sampling  often serves as a reweighting step within larger downstream algorithms, many of which are widely used in modern statistical workflows. To name a few, 
In Bayesian model evaluation, the \emph{leave-one-out} (LOO) predictive quantities are expectations under the  leave-one-out posterior \(p_i(\theta)\propto \pi(\theta\mid y_{-i})\).  Exact LOO requires refitting the model for deleting every data point. Instead,  importance-sampling-based LOO   \citep{gelfand1992model, peruggia1997variability, vehtari2017practical, magnusson2019bayesian, burkner2020approximate}  reuses the full posterior proposal \(q(\theta)\propto \pi(\theta\mid y)\) weighted by the inverse likelihood $r_i(\theta)=   \pi^{-1}(y_i | \theta)$. In  \emph{recursive Bayesian computing}  \citep{doucet2001introduction, chopin2002sequential, hooten2021making},  importance sampling is used to approximate expectations under a sequence of on-line targets. Draws from a previous distribution are reweighted by an incremental likelihood factor $r_i(\theta)= \pi(\theta\mid y_i)$ to approximate the updated posterior expectations.  In \emph{sensitivity analysis}, given posterior draws under one fixed model, prior, temperature, or dataset, importance sampling can approximate expectations under a tilted posterior \citep{smith1992bayesian, kallioinen2024detecting}. In \emph{approximate inference}, importance sampling  reweighs  variational samples for better calibration and diagnostics \citep{yao2018yes}. Conversely, using importance sampling inside variational objective tightens the evidence lower bound and leads to importance weighted variational inference \citep{burda2016importance,domke2018importance}.

Despite its wide deployment and asymptotic consistency, the weighted estimate \eqref{eq:IS} remains fragile \citep{vehtari2024pareto}. The usual central limit theorem 
requires a finite second moment of the weighted integrand, which often fails in practice. For example, even for a pair of Gaussian target and proposal,
$
p=\n(0,\sigma_p)
$
and
$
q=\n(0,\sigma_q),
$
then
$
\E_q[w^2]=\infty
$
as soon as $2\sigma_q^2\leq\sigma_p^2$. That is, under a modest scale mismatch, the $S^{-1/2}$ Monte Carlo convergence rate no longer applies to the IS estimate. In general, the required sample size $S$ grows exponentially in the target-proposal discrepancy \citep{chatterjee2018sample}; when this discrepancy scales linearly with parameter dimension $d$, the sample size $S$ is required at the order  $\exp(d)$. This curse of dimensionality rapidly makes direct importance weighting prohibitive in a moderate dimensional space.

\paragraph{Uncertainty quantification is challenging.}
Beyond its instability and potentially high or infinite variance, the weighted estimator \eqref{eq:IS} often lacks reliable uncertainty quantification.
The usual sample variance estimate \citep{geweke1989bayesian}  relies on the weighted empirical second moments:
\begin{equation}\label{eq:IS_var}
\widehat{\operatorname{Var}}(\hat\mu)
=
\begin{cases}
\displaystyle
\frac{1}{S^2}
\sum_{s=1}^S
\left (
w_s h(\theta_s)-\hat\mu_{\mathrm{IS}}
\right)^2,
& \text{normalized IS}, \\[1.4em]
\displaystyle
 
\frac{
 \sum_{s=1}^S
w_s^2
\left (
h(\theta_s)-\hat\mu_{\mathrm{SNIS}}
\right )^2
}{
\left( \sum_{s=1}^S w_s\right)^2
},
& \text{self-normalized IS}.
\end{cases}
\end{equation}
This variance estimate can be unreliable for three reasons: 
(i) The underlying population variance may not exist. When $\E_q[w^2]=\infty$, the variance of the IS estimate can be infinite even for bounded $h$, and 
the sample variance is meaningless.
(ii) Even when the variance is finite, when the importance ratios $w$ are heavy-tailed, this empirical variance is dominated by a small number of extreme weights $w$, even more severely than the mean as the variance involves the quadratic weight $w^2$. 
(iii) The self-normalized IS estimator is a ratio of dependent random variables and is $O(1/S)$ biased. The empirical variance does not fully assess the mean squared error.

The limitation of the sample variance has motivated several alternative IS diagnostics. Pareto smoothed importance sampling \citep[PSIS, ][]{vehtari2024pareto} instead reports the estimated Pareto tail index \(\hat k\) of the marginal distribution of $w$, while \citet{chatterjee2018sample} proposes to monitor $\max{w_s}/\sum w_s$. These diagnostics appear more stable than the classical effective sample size \citep[ESS,][]{kong1992note, elvira2022rethinking}, and are useful qualitative indicators of 
whether an IS estimate should be trusted at all, but they still do not provide a calibrated uncertainty statement or confidence interval for the IS estimate. For this difficulty,  the Monte Carlo uncertainty is often ignored or reduced to a diagnostic warning in many current statistical workflows involving importance sampling, such as model comparison, selection, and averaging \citep[e.g.,][]{vehtari2017practical, yao2018using},  despite the necessity to propagate it to downstream decision analysis.

\paragraph{Proposed method: Outline of poststratified importance sampling (PIS).}
We aim at the general importance sampling task: we can evaluate the continuous but potentially unnormalized target and proposal densities $p(\theta)$ and $q(\theta)$, and we have IID draws
$
\theta_{1}, \theta_2, \ldots, \theta_S \sim q(\theta).
$
Our new strategy is to continuously stratify the distribution by the scalar importance ratio
$
w=r(\theta).
$
Then we rearrange the target expectation using the law of total expectation 
\begin{equation}\label{eq_strata}
\E_{p}[h(\theta)]
=
\E_p\left[ \E_p[h(\theta)\mid w]\right]
=
\int_{0}^{\infty} p_w(w)\E_p[h(\theta)\mid w]\,dw.
\end{equation}
Here $\E_p[h(\theta)\mid w]$ denotes the conditional expectation given $r(\theta)=w$,  and $p_w(w)$ denotes the marginal density of the scalar random variable $w$, both under $\theta\sim p$; see Section~\ref{sec:math} for a formal notation.

We develop \emph{poststratified importance sampling} (PIS), a framework for computing $\E_{p}[h(\theta)]$  based on the identity \eqref{eq_strata}.  It follows the motivation of multilevel regression and poststratification \citep[MRP, ][]{gelman1997poststratification} in survey sampling. While MRP partitions a population into demographic strata and averages fitted stratum-level outcomes,  PIS uses the scalar importance ratio $w=r(\theta)$ as a continuous stratification variable. This representation admits a Rao-Blackwellization interpretation: we replace the original integrand  \(h(\theta)\) with the conditional expectation \(\E_p[h(\theta)\mid w]\), and then average over the marginal law \(p_w(w)\).

Both the (a)  conditional expectation and (b) marginal law are unknown and must be estimated. Different modeling choices for these two objects yield a family of PIS estimators by plugging in \eqref{eq_strata}, 
\begin{equation}\label{eq_PIS}
\hat \mu_{\text{PIS}}=  \int_{0}^{\infty} \hat p_w(w)\hat \E_p[h(\theta)\mid w]\,dw.    
\end{equation}
This general PIS framework recovers the classical importance weighting \eqref{eq:IS}  as a degenerate special  case. That is, if we adopt a ``model-free'' implementation, 
\begin{itemize}
    \item  estimate the conditional expectation using an unsmoothed estimate  $\hat \E_p[h(\theta)\mid w_i] \approx h(\theta_i),$
    \item   estimate the marginal law empirically $\hat q_w(w) \approx  S^{-1} \sum_{i=1}^S  \mathbf{1}(w=w_i)$.
\end{itemize}
For the equivalence that we will develop in Section~(\ref{sec:method}), this empirical $\hat q(w)$  implies  $\hat p(w) =   S^{-1} w_i \sum_{i=1}^S  \mathbf{1}(w=w_i)$ (when normalized) or $\hat p(w) \approx  {  \sum_{i=1}^S  w_i \mathbf{1}(w=w_i)}/{\sum_{i=1}^S  w_i} $ (when unnormalized). Thus, plugging-in this $\hat \E_p[h(\theta)\mid w]$ and  $\hat p(w)$, the PIS estimate $\hat \mu_{\text{PIS}}$ becomes the classical (self-normalized) importance weighting \eqref{eq:IS}.

Unsurprisingly, both $\E_p[h(\theta)\mid w]$ and $q_w(w)$ can admit more efficient estimates:
\begin{itemize}
    \item  The unsmoothed conditional mean estimate amounts to a degenerating, interpolating regression. It infers the conditional mean at input \(w_s\) using a single observation $h(\theta_i)$, without pooling information across samples.
\item The empirical estimate puts mass only on the observed weights, applies no tail smoothing, and discards additional information available from the density evaluation.  Its extrapolation can be especially poor when $q_w(w)$ is heavy-tailed.
\end{itemize}

These limitations motivate regularization and information
sharing. We implement PIS in three steps:
\begin{enumerate}
     \item \textbf{Conditional mean via one-dimensional regression.}  We fit a model-based regression to estimate the  conditional expectation $\E_p[h(\theta)\mid w_i]$  using $\theta$ samples from $q$, typically a Gaussian process regression.  This step uses the identity \(\E_p[h (\theta) \mid w]=\E_q[h(\theta) \mid w]\).
   \item  \textbf{Marginal law via one-dimensional  score-based heavy-tail estimation.} We estimate the marginal density \(p_w(w)\) and \(q_w(w)\) 
using a  flexible Dirichlet process mixture with a generalized Pareto tail for the distribution of \(w\). 
Our density estimation not only uses the sample \(w_i\), but also incorporates score evaluations of the densities \(p(\theta_i)\) and \(q(\theta_i)\), and this score information improves the tail estimation.
    \item  \textbf{Aggregation via one-dimensional quadrature.} After obtaining functional estimates of $\E_p[h(\theta)\mid w]$ and $p_w(w)$, we adopt one-dimensional numerical integration in \eqref{eq_PIS} to obtain the PIS estimate. 
\end{enumerate}

The modeling formulation also supports uncertainty
quantification and model checking. 
The modeling formulation also supports uncertainty quantification and model checking. Monte Carlo draws propagate uncertainty in the regression and marginal density estimates to the final PIS estimate, while posterior predictive checks can assess the adequacy of both models.

\subsection{Measure-theoretic background of the conditional and marginal laws}\label{sec:math} 
When \(p(\theta)\) and \(q(\theta)\)  are unnormalized,
$
Z_p=\int_{\mathbb R^d}p(\theta)\,d\theta
$
and 
$
Z_q=\int_{\mathbb R^d}q(\theta)\,d\theta
$
denotes the unknown normalizing constants. When we write \(\theta\sim p\) or \(\theta\sim q\), we mean that \(\theta\) is distributed from the normalized densities $p/Z_p$ or $q/Z_q$.  

Throughout this paper, when \(\theta\sim p\), we write \(p_w(w)\) for the marginal density of \(w=r(\theta)\), and \(p(\theta\mid w)\) for the conditional density of \(\theta\) given \(r(\theta)=w\). Likewise, \(q_w(w)\) and \(q(\theta\mid w)\) denote the corresponding marginal and conditional densities under \(\theta\sim q\). Since \(w\) is a deterministic function of \(\theta\), the formal definition of such marginal and conditional laws requires geometric measure theory \citep[e.g.,][]{evans2015measure}. Below, we briefly review this construction so that our notation is self-contained.

In this paper, we work with continuous distributions on \(\R^d\) equipped with the Borel \(\sigma\)-algebra and Lebesgue measure. Let \(p\) and \(q\) be two smooth, nonnegative, integrable functions on \(\mathbb R^d\), with finite normalizing constants \(Z_p,Z_q\) and full support on $\R^d$. Define the potentially unnormalized importance ratio function
$
w=r(\theta)=p(\theta) / q(\theta).
$
Given any value \(w\in[0,\infty)\), define the \(w\)-level set
\[
\Sigma_w
=
\{\theta\in\mathbb R^d:r(\theta)=w\}.
\]
We assume that, for almost every \(w\), the level set \(\Sigma_w\) is non-singular; that is, \(\nabla r(\theta)\neq 0\) for all \(\theta\in\Sigma_w\). This non-singularity makes \(\Sigma_w\) a smooth \((d-1)\)-dimensional submanifold.

Compared with regular importance sampling \eqref{eq:IS}, the additional assumptions are that (a) the importance ratio map \(r\) is smooth
, and (b) the level set is non-singular. The former assumption can fail, for example, when using a step-function proposal to approximate a normal density. The latter assumption can fail, for example, when the importance ratio \(r(\cdot)\) is constant. Other than such edge cases, these mild conditions hold in nearly all importance sampling applications we consider.

\emph{Marginal law.}
Let \(P_W\) denote the pushforward of the probability measure \(\{p(\theta)/Z_p\}d\theta\) under the map \(r\). That is, for every Borel set \(A\subset[0,\infty)\),
\[
P_W(A)
=
\int_{\{\theta:\,r(\theta)\in A\}}
\frac{p(\theta)}{Z_p}\,d\theta.
\]

\emph{Conditional expectation.}
For any integrable function \(h\), a version of the conditional expectation of \(h(\theta)\) given \(r(\theta)=w\) under \(p\) is a measurable function of \(w\), satisfying, for every Borel set \(A\subset[0,\infty)\),
\begin{equation}\label{eq:conditionalexp}
\int_{\{\theta:\,r(\theta)\in A\}}
h(\theta)\frac{p(\theta)}{Z_p}\,d\theta
=
\int_A \E_p[h(\theta)\mid r(\theta)=w] \,P_W(dw).
\end{equation}
Taking \(A=[0,\infty)\) in \eqref{eq:conditionalexp} gives the total-expectation formula
\[
\E_p[h(\theta)]
=
\int_{0}^{\infty} \E_p[h(\theta)\mid r(\theta)=w] \,P_W(dw).
\]
Its validity follows directly from the definition; no application of Fubini's theorem is required.

Further define \(P_p(d\theta\mid w)\), a version of the conditional probability measure of \(\theta\) given \(r(\theta)=w\) under \(p\), supported on \(\Sigma_w\), such that for any integrable function \(h\),
\[
\E_p[h(\theta)\mid r(\theta)=w]
=
\int_{\Sigma_w}
h(\theta)\,
P_p(d\theta\mid r(\theta)=w).
\]

\emph{Surface measure.}
One key mathematical device for the score-based estimator in this paper is the coarea formula, which turns the abstract conditional measure on the level set into an explicit surface-measure representation. Let \(d\sigma_w(\theta)\) denote the \((d-1)\)-dimensional surface measure on \(\Sigma_w\). Then, for any Borel sets \(A\subset[0,\infty)\) and \(B\subset\mathbb R^d\), the coarea formula gives
\begin{equation}\label{eq:coarea_sets}
\int_{\{\theta:\,r(\theta)\in A,\ \theta\in B\}}
\frac{p(\theta)}{Z_p}\,d\theta
=
\int_A
\left[
\int_{\Sigma_w\cap B}
\frac{p(\theta)}{Z_p}
\frac{1}{\|\nabla r(\theta)\|}
\,d\sigma_w(\theta)
\right]dw.
\end{equation}

As a consequence, when \(P_W\) admits a density with respect to Lebesgue measure on \(\mathbb R\), denoted by \(p_w(w)\), we have
\begin{equation}\label{eq:pW_coarea}
p_w(w)
=
\int_{\Sigma_w}
\frac{p(\theta)}{Z_p}
\frac{1}{\|\nabla r(\theta)\|}
\,d\sigma_w(\theta).
\end{equation}
The conditional density of $\theta$ given $r(\theta)=w$
 with respect to the surface measure is
\begin{equation}\label{eq:conditional_coarea}
p(\theta\mid r(\theta)=w)
=
\frac{
p(\theta)\|\nabla r(\theta)\|^{-1}
}{
\int_{\Sigma_w}
p(\tilde\theta)\|\nabla r(\tilde\theta)\|^{-1}
\,d\sigma_w(\tilde\theta)
},
\qquad \theta\in\Sigma_w.
\end{equation}

In this paper, we reserve \(w\) to denote the importance weight. When there is no ambiguity, we write
$p(\theta\mid w)$
and
$\E_p[h(\theta)\mid w]$
as shorthand for
$p(\theta\mid r(\theta)=w)$
and
$\E_p[h(\theta)\mid r(\theta)=w]$
respectively. The analogous notation \(q(\theta\mid w)\) and \(\E_q[h(\theta)\mid w]\) is used under the proposal distribution \(q\).

\section{Proposed  method}\label{sec:method}
We propose poststratified importance sampling (PIS).  It implements the conditioning strategy  in \eqref{eq_PIS}, and consists of three steps: (i) estimation of the conditional expectation, (ii) marginal density, and (iii) numerical integral. Below, we  develop a general framework that reformulates importance sampling into learning tasks and provides a practical recipe for these three steps.

\subsection{Conditional expectation via regression}\label{sec:method_reg} 
The first step is to estimate $\E_p[h(\theta)\mid w]$, the conditional expectation under the target $p$. To emphasize that it is a function of $w$,  we write 
$$m(w)\coloneqq  \E_p[h(\theta)\mid w].$$
The observed pairs $(w_s,h(\theta_s))$ suggest estimating
the conditional mean by regression. Although the draws come
from the proposal, the following theorem shows that the
conditional mean is the same under $p$ and $q$, so the
regression requires no importance weighting.

\begin{theorem}\label{prop:conditional_equivalence}
Let $p$ and $q$ be probability densities on $\mathbb R^d$ with full
support, and let $W=p(\theta)/q(\theta)$. Then:
\begin{enumerate}
\renewcommand{\labelenumi}{(\theenumi)}

    \item For every integrable measurable function $h$,
    \begin{equation}
    \E_p[h(\theta)\mid W=w] = \E_q[h(\theta)\mid W=w].
    \end{equation}

    \item For every integrable function $h$, suppose that a map $T=T(\theta)$ satisfies
    \begin{equation}\label{eq:constant_mean}
    \E_p[h(\theta)\mid T=t]
    =
    \E_q[h(\theta)\mid T=t]
    \end{equation}
    Then $T$ determines $W$: there exists a measurable function $\alpha$ such that $W=\alpha(T(\theta))$ almost surely.

\end{enumerate}
\end{theorem}

With the proposal sample $\theta_s \sim q$, we obtain training pairs
$$
\mathcal{D}_q=\{(h_s \coloneqq h(\theta_s), w_s \coloneqq r(\theta_s)) \mid 1\leq s\leq S \},$$ on which we fit a Gaussian process (GP) regression, with, for example, a squared exponential kernel: 
$$h_{i} \mid m \sim \mbox {normal} (m(w_i), \sigma); \quad 
m(w) \sim  \mathcal{GP}(0, K(w,w')),  \quad K(w, w^\prime) = \tau^2 \exp\left(-\frac{|w-w^\prime|^2}{\rho^2}\right).
$$
This basic GP model can be readily extended when suggested by the data.  For example, we can 
accommodate heteroskedasticity by modeling residual scale, $\sigma=\sigma(w)$.

When the integrand \(h(\theta)\) is not continuous, it is straightforward to go beyond the Gaussian likelihood. For example,  when $h(\theta)$ is binary, we use a
Bernoulli-likelihood GP to estimate $\Pr(h(\theta) =1 \mid  w)$. 

More generally, the regression can be performed on any bijective transformation of the weight, \(\psi(w)\), as conditioning on \(w\) is equivalent to conditioning on \(\psi(w)\). The transformation \(\psi(w)\) becomes a model choice and can, in principle, be informed by the data. In practice, for numerical stability, we often use the log-weight scale, \(\psi(w)=\log w\), in this Gaussian process regression.

Given evaluation points $(\tilde w_1,\ldots,\tilde w_L)$, the fitted GP provides posterior means, covariances, and joint posterior draws of the conditional mean function $
(m(\tilde w_1),\ldots,m(\tilde w_L))\mid \mathcal D_q.
$ We use these
draws to propagate regression uncertainty through the
final PIS estimator.



\subsection{Marginal density estimation with a Pareto-tailed Dirichlet process mixture}\label{sec:method_den} 
The second step is to estimate $p_w(w)$, the marginal density of $w$ under $p$. Although $w$ is one-dimensional, this density estimation can be challenging for two reasons:
\begin{enumerate}[label=(\alph*)]
    \item Available draws of $w_i= r(\theta_i)$ are under the proposal $\theta\sim q$, not the target. 
     \item This $p_w(w)$ has support on $[0, \infty)$, and typically contains a heavy right-tail.
\end{enumerate}

To solve problem (a), we show that the only difference between $p_w(w)$ and $q_w(w)$, the marginal density of $w=r(\theta)$ under $p$ and $q$, is a multiplicative factor $w$:

\begin{theorem}\label{prop:marginal_weight_law}

Let $W=p(\theta)/q(\theta)$, and let $p_w$ and $q_w$ denote its marginal densities under the corresponding normalized target and proposal distributions, respectively

\begin{enumerate}
    \item If both $p(\theta)$ and $q(\theta)$ are normalized, the marginal densities of $W$ under $p$ and $q$ satisfy, for a.e. $w$,
    $$
    p_w(w)=wq_w(w).
    $$
    
    \item If either $p(\theta)$ or $q(\theta)$ is not normalized, the marginal densities of $W$ under their normalized versions satisfy, for a.e. $w$,
    $$
    p_w(w)\propto wq_w(w).
    $$
\end{enumerate}

\end{theorem}

Therefore, estimating one proposal marginal density function  $q_w$  immediately determines the target $p_w$  (or up to self-normalizing). This is a deterministic relation between two functions $p_w(\cdot)$ and $q_w(\cdot)$, not an additional importance weighting. 

To address the challenge (b),
we develop a model-based density estimate of $q_w(\cdot)$ with two features: one on the parametric form and one on the inference:
\begin{itemize}
    \item \emph{Model-based regularization using a heavy-tailed density}: we estimate $q_w(\cdot)$ using a Dirichlet process mixture of truncated-normal densities for the bulk and a generalized Pareto model for the tail.
    \item \emph{Score-based tail learning}:  We estimate the Pareto tail index using not only the observed sample but also local score information.
\end{itemize}

\paragraph{Flexible bulk–tail density model.}
We construct a flexible one-dimensional model for $q_w$ by combining
a Dirichlet process mixture (DPM) for the bulk with a generalized
Pareto distribution (GPD) for the right tail.
Following the bulk-tail separation in PSIS \citep{vehtari2024pareto},
we treat the largest
$M=\lfloor\min(3\sqrt{S},\,0.2S)\rfloor$ observed weights as the tail.
The threshold and tail fraction are
\[
w_0 \coloneqq w_{(S-M)},
\qquad
\pi \coloneqq M/S.
\]

We model the bulk as a mixture of normal densities truncated
to $[0,w_0]$:
\[
q_{\mathrm{bulk}}(w\mid G)
=
\int \phi_{[0,w_0]}^{\mu,\sigma}(w)\,
G(\mathrm d\mu,\mathrm d\sigma),
\qquad
G\sim\operatorname{DP}(\alpha G_0),
\]
where $\phi_{[0,w_0]}^{\mu,\sigma}$ denotes the truncated-normal
density and $\alpha>0$ is the concentration parameter.
We specify the base distribution $G_0$ over component means
and scales through the priors:
\[
\mu\sim \n(\mu_0,s_\mu)\mid 0\leq\mu\leq w_0,
\qquad
\sigma^2\sim\operatorname{InvGamma}(a_\sigma,b_\sigma)
\mid 0<\sigma^2\leq\overline{\sigma}^2.
\]

We model the tail  using a generalized
Pareto distribution (GPD) with a shape parameter $k$ and a scale $\tau>0$,
\[
q_{\mathrm{GPD}}^{k,\tau}(w)
=
\frac{1}{\tau}
\left(1+\frac{k(w-w_0)}{\tau}\right)^{-1-1/k},
\qquad w>w_0.
\]

Combining the bulk and tail gives the marginal density 
\begin{equation}\label{def:tn-dpm-gpd}
q_w(w\mid G,k,\tau)
=
(1-\pi)q_{\mathrm{bulk}}(w\mid G)
+
\pi q_{\mathrm{GPD}}^{k,\tau}(w).
\end{equation}
The unknown quantities in the marginal model are the bulk mixing distribution $G$
and the tail Pareto parameters $(k,\tau)$. We infer $G$ from the bulk
observations $w_s\leq w_0$ using standard Bayesian inference
via Markov chain Monte Carlo (MCMC). To stabilize estimation of the tail index $k$, we incorporate score information when fitting $(k,\tau)$ to the
tail observations $w_s>w_0$, as described below.


\begin{figure}[h!]
\centering
\begin{tikzpicture}[
    x=1.25cm,
    y=12cm,
    font=\normalsize
]

\path[fill=pisgreen!10,draw=none]
(0,0)
-- plot[
    domain=0:4,
    samples=150,
    smooth,
    variable=\x
]
({\x},{0.486*\x*exp(-0.7*\x)})
-- (4,0)
-- cycle;

\path[fill=pisred!10,draw=none]
(4,0)
-- plot[
    domain=4:10,
    samples=150,
    smooth,
    variable=\x
]
({\x},{
    0.486*4*exp(-0.7*4)
    *pow(1+0.25*(\x-4),-3)
})
-- (10,0)
-- cycle;

\draw[pisgreen,very thick]
plot[
    domain=0:4,
    samples=150,
    smooth,
    variable=\x
]
({\x},{0.486*\x*exp(-0.7*\x)});

\draw[pisred,very thick]
plot[
    domain=4:10,
    samples=150,
    smooth,
    variable=\x
]
({\x},{
    0.486*4*exp(-0.7*4)
    *pow(1+0.25*(\x-4),-3)
});

\draw[densely dashed,gray]
(4,0) -- (4,0.285);

\draw[-{Latex[length=2.7mm,width=1.8mm]},thick]
(0,0) -- (10.8,0)
node[below right] {$w$};

\draw[-{Latex[length=2.7mm,width=1.8mm]},thick]
(0,0) -- (0,0.34)
node[above] {$q_w(w)$};

\draw (0.7,0.004) -- (0.7,-0.004)
node[below,font=\small] {$w_{(1)}$};

\draw (2.2,0.004) -- (2.2,-0.004)
node[below,font=\small] {$\cdots$};

\draw (4,0.004) -- (4,-0.004)
node[below,font=\small] {$w_{(S-M)}$};

\draw (5.7,0.004) -- (5.7,-0.004)
node[below,font=\small] {$w_{(S-M+1)}$};

\draw (7.6,0.004) -- (7.6,-0.004)
node[below,font=\small] {$\cdots$};

\draw (9.7,0.004) -- (9.7,-0.004)
node[below,font=\small] {$w_{(S)}$};

\node[pisgreen,align=center] at (1.65,0.295)
{\textbf{DPM bulk}};

\node[pisred,align=center] at (7.0,0.125)
{\textbf{GPD tail}};

\node[pisgreen!80!black,font=\large] at (1.7,0.04)
{$1-\pi$};

\node[pisred!80!black,font=\large] at (4.7,0.04)
{$\pi$};

\end{tikzpicture}

\caption{ $q_w(w)$: the marginal distribution of the scaler importance ratio under the proposal. We model the tail area (largest $M$ samples) by a generalized Pareto distribution, and the remaining bulk by a Dirichlet process mixture.}
\label{fig:dpm-gpd-density}
\end{figure}
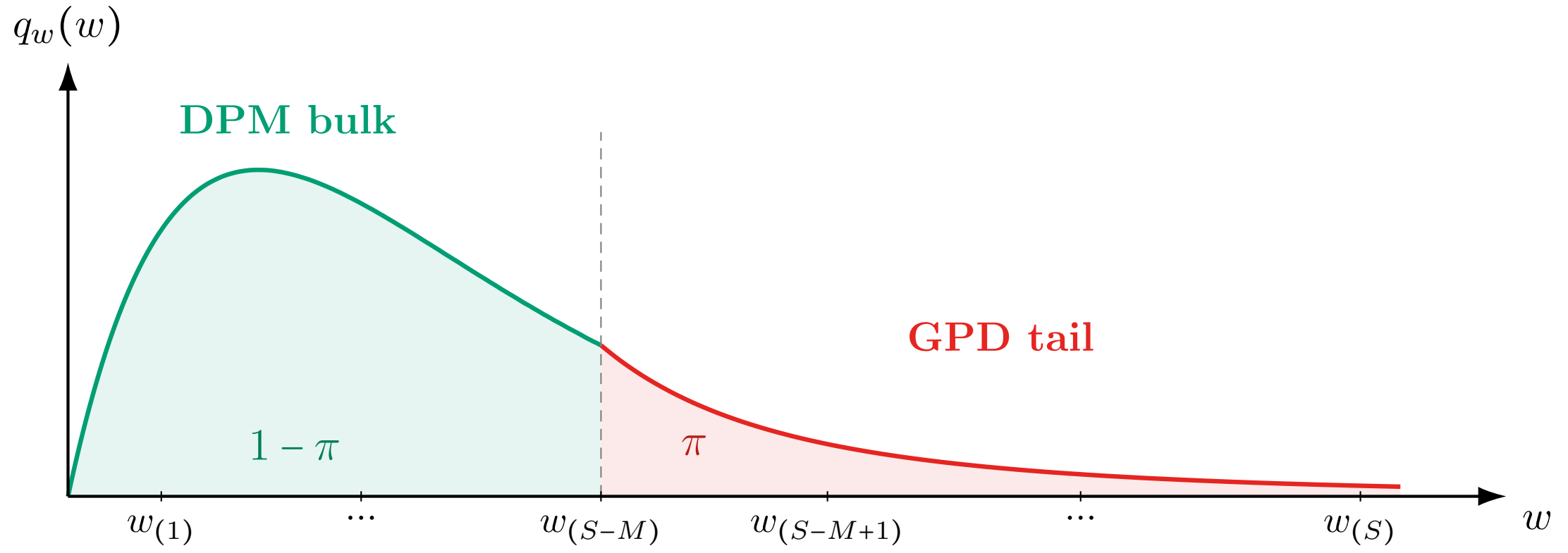

\paragraph{Score-based tail estimation.}
We can estimate the GPD parameters from the observed tail
weights ($w_{(S-M)}, w_{(S-M+1)}, \ldots, w_{S}$) using standard likelihood-based inference, as in heavy-tailed
density  estimation \citep{hickling2024flexible} and PSIS
\citep{vehtari2024pareto}. For example, a Bayesian fit gives posterior density 
\begin{equation}\label{eq_lik_based}
p^{\mathrm{post}}(k,\tau | w_{(S-M)}, w_{(S-M+1)}, \ldots, w_{(S)})
\propto
\mathrm{prior}(k,\tau)
\prod_{j=1}^{M}
q_{\mathrm{GPD}}^{k,\tau}\left(w_{(S-M+j)}\right).
\end{equation}
Optionally, when both densities $p$ and $q$ are differentiable,  we can further improve the tail index fit by using derivative information. Since
$w=r(\theta)$ with $\theta\sim q$, the marginal score of
$q_w$ admits the following conditional expectation
representation, which enables a score-based
estimator.
\begin{lemma}\label{thm:score}
Assume that $p$ and $q$ are smooth density functions on $\R^d$. Let $q_w(w)$ denote the marginal density of $w=p(\theta)/q(\theta)$ under $q$. Fix an open interval $I\subset\mathbb R$. Assume that
$\|\nabla r(\theta)\|>0$ for all  $\theta\in r^{-1}(I)$.
Then, for every $w\in I$ such that $q_w(w)>0$,
$$
\nabla_w \log q_w(w)
=
\mathbb E_q\left[s(\theta)\mid r(\theta)=w\right],
$$
where
$$
s(\theta)
:=
\frac{\nabla r(\theta)\cdot \nabla \log q(\theta)}
{\|\nabla r(\theta)\|^2}
+
\operatorname{div}\left(
\frac{\nabla r(\theta)}
{\|\nabla r(\theta)\|^2}
\right).
$$
\end{lemma}
Since $s(\theta_s)$ is unbiased for the marginal
score at $w_s$, matching these estimates to the GPD score over
the largest $M$ weights gives a score matching estimate 
\[
(\hat k,\hat\tau)
=
\arg\min_{k,\tau}
\sum_{j=1}^{M}
\left[
s\!\left(\theta_{(S-M+j)}\right)
+
\frac{1+k}{\tau+k\left(w_{(S-M+j)}-w_0\right)}
\right]^2.
\]
Incorporating derivatives of $p$ and $q$ provides additional information beyond the observed weights and can improve tail-index estimation relative to likelihood-based fitting.  A companion paper develops the general theory of this
score-based approach.

\begin{algorithm}[t]
\caption{Score based marginal density estimation of $w$}
\label{alg:score_marginal_density}
\begin{algorithmic}[1]

\State \textbf{Input:} Proposal draws
$\theta_1,\ldots,\theta_S\sim q$, importance ratio $w=p/q$,
unbiased score expression $s(\theta)$, tail size $M$, and  output Monte Carlo size $B$.
\State \textbf{Output:} Estimated density $\hat q_w$ and
posterior density draws $q_w^{(1:B)}$.
\State Set $w_i\gets p(\theta_i)/q(\theta_i)$ and $s_i\gets s(\theta_i)$
for $i=1,\ldots,S$.
\State Sort the pairs $(w_i,s_i)$ so that
$w_{(1)}\leq\cdots\leq w_{(S)}$.

\State Set $w_0\gets w_{(S-M)}$ and $\pi\gets M/S$.

\State Estimate the generalized Pareto shape and scale by score matching:
$$
(\hat k,\hat\tau)
\gets
\arg\min_{k,\tau}
\sum_{i=1}^{M}
\left[
s_{(S-i+1)}
-
\left.
\frac{\mathrm d}{\mathrm dw}
\log q_{\mathrm{GPD}}^{k,\tau}(w)
\right|_{w=w_{(S-i+1)}}
\right]^2.
$$

\State Set
$$
\hat q_{\mathrm{tail}}(w)
\gets
q_{\mathrm{GPD}}^{\hat k,\hat\tau}(w-w_0)
\mathbbm 1\{w>w_0\}.
$$

\State Fit a Dirichlet process mixture to
$w_{(1)},\ldots,w_{(S-M)}$ on $[0,w_0]$, obtain its posterior mean
density $\hat q_{\mathrm{bulk}}$, and retain posterior draws
$G^{(1)},\ldots,G^{(B)}$.

\State Set
$$
\hat q_w(w)
\gets
(1-\pi)\hat q_{\mathrm{bulk}}(w)
+
\pi\hat q_{\mathrm{tail}}(w).
$$

\State For $b=1,\ldots,B$, set
$$
q_w^{(b)}(w)
\gets
(1-\pi)q_{\mathrm{bulk}}(w\mid G^{(b)})
+
\pi\hat q_{\mathrm{tail}}(w).
$$

\State \Return $\hat q_w$ and $q_w^{(1:B)}$.

\end{algorithmic}
\end{algorithm}


\subsection{Numerical integration}\label{sec:method_int} 
So far we have constructed the functional estimate of  $m(w)$ and $q_w(w)$, and $p_w(w)$ is a deterministic transformation of $q_w(w)$. 
The final step is to compute the one-dimensional integral
\begin{equation}\label{eq:inergral}
    \mu \coloneqq
\E_p[h(\theta)]
=
\int_{0}^\infty \hat m(w) \hat p_w(w)\,dw
\end{equation}
We construct a quadrature on a $w$-grid
$
0<w_1^\star<\cdots<w_L^\star.
$

\paragraph{PIS estimate when $p$ and $q$ are normalized.}
Theorem~\ref{prop:marginal_weight_law} translates the proposal marginal $\hat q_w(\cdot)$ into the functional  estimation of the target marginal: 
$$
\hat p_w(w)=\hat q_w(w) w.
$$
Using the trapezoidal rule in \eqref{eq:inergral}, the target expectation becomes 
\begin{align}
\hat\mu_{\mathrm{PIS}}
&=
\sum_{\ell=1}^{L-1}
\frac{w_{\ell+1}^\star-w_\ell^\star}{2}
\left(
\hat m(w_\ell^\star)\hat p_w(w_\ell^\star)
+
\hat m(w_{\ell+1}^\star)\hat p_w(w_{\ell+1}^\star)
\right)\nonumber 
\\  
&=\sum_{\ell=1}^{L-1}
\frac{w_{\ell+1}^\star-w_\ell^\star}{2}
\left(
w_\ell^\star \hat m(w_\ell^\star)\hat q_w(w_\ell^\star)
+
w_{\ell+1}^\star \hat m(w_{\ell+1}^\star)\hat q_w(w_{\ell+1}^\star)
\right).\label{eq:psis_trapezoid}
\end{align} 
 
\paragraph{Self-normalized PIS estimate density when $p$ and $q$ are unnormalized.}
When only unnormalized densities are available, $w$ represents the unnormalized density ratio. The target marginal density requires self-normalization. 
\begin{equation}\label{eq:selfnormalized_p}
    \hat p_{W}(w)
=
\frac{w\hat q_{W}(w)}
{\int_0^\infty u\hat q_{\tilde W}(u)\,du}.
\end{equation}

We compute both the integral \eqref{eq:inergral} and the self-normalization factor in \eqref{eq:selfnormalized_p} with two one-dimensional quadratures
\begin{align}
\hat\mu_{\mathrm{PIS}}
&=
\frac{
\sum_{\ell=1}^{L-1}
\frac{w_{\ell+1}^\star-w_\ell^\star}{2}
\left(
w_\ell^\star \hat m(w_\ell^\star)\hat q_w(w_\ell^\star)
+
w_{\ell+1}^\star \hat m(w_{\ell+1}^\star)\hat q_w(w_{\ell+1}^\star)
\right)}{{ \sum_{\ell=1}^{L-1}
\frac{w_{\ell+1}^\star-w_\ell^\star}{2}
\left(
w_\ell^\star \hat q_w(w_\ell^\star)
+
w_{\ell+1}^\star \hat q_w(w_{\ell+1}^\star)
\right)}}.
\label{eq:psis_trapezoid_unnormalized}
\end{align}

\subsection{Uncertainty quantification using posterior draws}

\begin{algorithm}[t]
\caption{Poststratified importance sampling}
\label{alg:PIS}
\begin{algorithmic}[1]

\State \textbf{Input:} Proposal draws $\theta_{1:S}\sim q$,
integrand $h$, score $s$, and tail size $M$.

\State \textbf{Output:} PIS estimate $\hat\mu_{\mathrm{PIS}}$ and
posterior uncertainty draws.

\State Initialize the quadrature grid
$0=w_1^\star<\cdots<w_L^\star=T$.

\State Set $w_s\gets p(\theta_s)/q(\theta_s)$ and $h_s\gets h(\theta_s)$
for $s=1,\ldots,S$.

\State Fit a Gaussian process to $\{(w_s,h_s)\}_{s=1}^S$ and evaluate
$\hat m_\ell\gets\hat m(w_\ell^\star)$ for $\ell=1,\ldots,L$.

\State Apply Algorithm~\ref{alg:score_marginal_density} to obtain
$\hat q_w$ and posterior density draws $q_w^{(1)},\ldots,q_w^{(B)}$.

\State Evaluate $\hat q_\ell\gets\hat q_w(w_\ell^\star)$ and set
$\hat p_\ell\gets w_\ell^\star\hat q_\ell$ for $\ell=1,\ldots,L$.

\If{$p$ and $q$ are normalized}
\State Set
$$
\hat\mu_{\mathrm{PIS}}
\gets
\sum_{\ell=1}^{L-1}
\frac{w_{\ell+1}^\star-w_\ell^\star}{2}
\left(
\hat m_\ell\hat p_\ell
+
\hat m_{\ell+1}\hat p_{\ell+1}
\right).
$$
\Else
\State Set
$$
\hat\mu_{\mathrm{PIS}}
\gets
\frac{
\displaystyle
\sum_{\ell=1}^{L-1}
\frac{w_{\ell+1}^\star-w_\ell^\star}{2}
\left(
\hat m_\ell\hat p_\ell
+
\hat m_{\ell+1}\hat p_{\ell+1}
\right)
}{
\displaystyle
\sum_{\ell=1}^{L-1}
\frac{w_{\ell+1}^\star-w_\ell^\star}{2}
\left(
\hat p_\ell+\hat p_{\ell+1}
\right)
}.
$$
\EndIf

\State Draw
$\bigl(m^{(b)}(w_1^\star),\ldots,m^{(b)}(w_L^\star)\bigr)$
jointly from the Gaussian process posterior for $b=1,\ldots,B$.

\State Apply the same quadrature rule, with the same normalization
when required, to each pair $(m^{(b)},q_w^{(b)})$ to obtain
$\hat\mu_{\mathrm{PIS}}^{(1)},\ldots,\hat\mu_{\mathrm{PIS}}^{(B)}$.

\State \Return $\hat\mu_{\mathrm{PIS}}$ and
$\hat\mu_{\mathrm{PIS}}^{(1:B)}$.

\end{algorithmic}
\end{algorithm}

The fitted conditional mean and marginal density carry uncertainty from the finite proposal sample, analogous to Monte Carlo error in ordinary importance sampling.
On a fixed quadrature grid $0<w_1^\star<\cdots<w_L^\star$,
the formulas \eqref{eq:psis_trapezoid} and
\eqref{eq:psis_trapezoid_unnormalized} are deterministic functions of the conditional means and marginal densities
on the grid. The point estimate plugs in the posterior means $\hat m$ and $\hat q_w$; to propagate their uncertainty,
we instead apply the same quadrature to posterior draws of these functions. 
For each $b=1,\ldots,B$, let
\[
\left\{
m^{(b)}(w_\ell^\star),
q_w^{(b)}(w_\ell^\star)
\right\}_{\ell=1}^L
\]
denote joint draws of the conditional mean and proposal marginal density on the grid. We draw $m^{(b)}$ jointly
across the grid from the GP posterior for the latent mean function and construct $q_w^{(b)}$ from posterior draws
of the DPM bulk and the fitted GPD tail.

For normalized $p$ and $q$, each pair gives
\[
\mu^{(b)}
=
\sum_{\ell=1}^{L-1}
\frac{w_{\ell+1}^\star-w_\ell^\star}{2}
\left(
w_\ell^\star m^{(b)}(w_\ell^\star)q_w^{(b)}(w_\ell^\star)
+
w_{\ell+1}^\star
m^{(b)}(w_{\ell+1}^\star)q_w^{(b)}(w_{\ell+1}^\star)
\right).
\]
The resulting draws $\mu^{(1)},\ldots,\mu^{(B)}$ approximate
the model-based posterior distribution of the target expectation.  For unnormalized densities, we apply the self-normalized
quadrature formula \eqref{eq:psis_trapezoid_unnormalized}
to each pair of function draws.


\subsection{Computational complexity}
At first glance, our model-based PIS appears to replace a Monte Carlo mean with a conceptually more elaborate formulation involving (a) Gaussian process regression, (b) density estimation, and (c) numerical quadrature. In fact, all three components are one-dimensional, yielding a scalable, nearly dimension-free computational cost.  
In particular, because both the outcome $h_s$ and input $w_s$ are scalars, the regression is independent of the parameter dimension $d$. One-dimensional fast  GP algorithms \citep[e.g., ][]{yadav2021faster, greengard2025efficient, greengard2025equispaced} are highly scalable,  achieving nearly linear computational cost in the sample size $S$.

The computational cost of the one dimensional quadrature is linear in the grid size $L$, and the trapezoidal quadrature error drops at the rate $O(L^{-2})$.  We therefore do not further optimize the quadrature scheme in this paper, although we expect that more sophisticated quadrature methods will further improve accuracy.

Moreover, PIS is not tied to a particular modeling choice. Our general framework provides a spectrum of complexity-accuracy tradeoffs through different choices of regression and density estimators, ranging from a simple empirical estimate to deep Bayesian models. 
Consequently, PIS bridges the gap between two extremes: a one-shot importance-sampling correction and full resampling from the target distribution, trading a small one-dimensional computation for improved accuracy and uncertainty quantification.
In many applications we consider, including cross-validation, sensitivity analysis, and approximate Bayesian inference, the dominant cost is obtaining the proposal sample, while the extra cost of 
fitting a one-dimensional regression and density model is only a fraction. 

\section{Theoretical analysis}\label{Sec:theory}
 
\subsection{Oracle properties}
We first analyze the idealized setting in which the conditional mean $m(w)=\E_p[h(\theta)\mid r(\theta)=w]$
and the marginal law $p_w(w)$ are known exactly.  We study two sources
of improvement in PIS: conditioning removes the conditional variation in \(\theta\mid r(\theta)=w\), while numerical quadrature removes the Monte Carlo variation associated with sampling over \(w\).

\paragraph{Oracle gain from conditioning.}
Suppose first that $p$ and $q$ are normalized. Ordinary importance sampling extrapolates from $q(\theta)$ to $p(\theta)$ in a potentially high-dimensional parameter space: $
\E_p[h(\theta)]
=
\E_q[wh(\theta)]
$. Instead, when $m(w)$ is known, PIS extrapolates from $q_w$ to $p_w$ in the one-dimensional weight space:
$\E_p[h(\theta)]
=
\int m(w)p_w(w)\,dw
=
\int m(w)wq_w(w)\,dw$.


To separate the gains from conditioning and quadrature, we introduce a quadrature-free
Monte Carlo version of PIS for theoretical purposes:
\begin{equation}
\mathbb E_p[h(\theta)]
\approx
\hat\mu_{\mathrm{MC\text{-}PIS}}
:=
\frac{1}{S}
\sum_{s=1}^S
w_s\hat m(w_s),
\end{equation}
This is equivalent to performing importance sampling in the \emph{one-dimensional} \(w\)-space, with the proposal $q_w$ and target $p_w$, and remains unbiased. 
When  $\E_q[w^2h(\theta)^2]<\infty$, 
direct application of the  law of total variance yields a non-negative variance reduction from the ordinary \(\theta\)-space  IS to our oracle \(w\)-space IS:
\[
\operatorname{Var}_q\!\left(\mu_{\mathrm{IS}} \right)
-
\operatorname{Var}_q\!\left(\mu_{\mathrm{MC\text{-}IS}}\right)
=
\frac{1}{S}\E_q\!\left[
w^2\operatorname{Var}_q\!\left(h(\theta)\mid w\right)
\right],
\]

\paragraph{Examples.}
To demonstrate the gain from conditioning, first, consider an importance sampling task with \(d\)-dimensional multivariate normal  proposal and target,
\[
p(\theta) =\mbox{MVN}(\theta \mid 0,I_d),
\quad
q(\theta)=\mbox{MVN}(\theta \mid u,I_d),
\quad
h(\theta)=a^\top\theta.
\]
where \(u\in\mathbb R^d\) is a fixed mean mismatch and $h(\theta)=a^\top\theta$ is a given projection of interest.  The oracle mean function is linear in $\log w$: 
$
m(w)
=
\frac{a^\top u}{\|u\|^2}
\left(
\frac{\|u\|^2}{2}-\log w
\right).$
The gain from conditioning depends on the alignment between the projection direction $a$ and the mismatch $u$. When $a=u$, $h(\theta)$ is determined
by $w$, and conditioning gives no variance reduction. If $a$ is orthogonal
to $u$, then $m(w)=0$, and the oracle PIS estimator has zero variance.

In general, suppose $a$ is uniformly distributed on the unit sphere. Averaging over projection directions, the variance
reduction is
\[\mathbb E_a\mathbb E_q\!\left[
w(\theta)^2\operatorname{Var}_q(a^\top\theta\mid w)
\right]
=
\mathbb E_q[w(\theta)^2]\,
\mathbb E_a\!\left(
\|a\|^2-\frac{(a^\top u)^2}{\|u\|^2}
\right)
=
\left(1-\frac{1}{d}\right)\exp(\|u\|^2).
\]
Averaged over all directions, the ratio of the expected variance of IS and oracle PIS is  
\[
\frac{\E_a[\Var(\mu_{\mathrm{IS}} )]}
{\E_a[\Var(\mu_{\mathrm{MC\text{-}PIS}} )]}
=
\frac{d+\|u\|^2}{1+\|u\|^2}.
\]
For a fixed $\|u\|$,  this variance reduction ratio grows linearly with the dimension $d$.

Next, to give an even sharper contrast, consider importance sampling under a shape-mismatch between a multivariate Student-\(t\) and normal distributions:
\[
p=t_\nu(0,I_d),
\quad
q=\mbox{MVN}(0,I_d),
\quad
h(\theta)=a^\top\theta,
\]
with $\nu>1$, and 
$\|a\|=1$.
Elementary calculations show that the mean function is a constant 
$
m(w)=\E_q[a^\top\theta\mid w]=0.
$
Hence, regardless of the direction $a$ or the dimension, conditioning removes the entire variance
$$
\Var(\hat\mu_{\mathrm{MC\text{-}PIS}})= S^{-1}\Var_q(wm(w)) =0
 $$
For every finite $\nu>1$, however, the ordinary IS summand
has infinite variance:
$$
\Var_q(w a^\top\theta)=\infty.
$$


\paragraph{Oracle gain from quadrature.}
Beyond conditioning, PIS can further reduce integration error by replacing the Monte Carlo average with one-dimensional quadrature. Standard trapezoidal error bounds apply to smooth integrands on compact intervals, whereas PIS integrates over $[0,\infty)$ with potentially heavy-tailed
weights. The following lemma extends the standard error bound to our setting.

\begin{lemma}[Oracle quadrature error]
\label{lem:oracle-quadrature}
Suppose $p$ and $q$ are normalized, and $m$ and $q_w$ are known. Assume that $q_w$ has a generalized Pareto
tail above a fixed threshold $w_0>0$, with shape $0<k<1$. Suppose
$m p_w\in C^2([0,w_0])$ and
$m\in C^2((w_0,\infty))$, with
\[
|m^{(j)}(w)|\leq Cw^{\beta-j},
\qquad j=0,1,2,\qquad w>w_0,
\]
for some $C<\infty$ and $\beta<(1-k)/k$.
Then using  $L$ quadrature nodes and a tail-adapted composite trapezoidal rule (see Appendix~\ref{sec_app_Quadrature}) yields a quadrature  error rate
\begin{equation}\label{eq:quadrature_error}
\left|\hat\mu_{\mathrm{PIS}}-\mu\right|=O(L^{-2}).    
\end{equation}
\end{lemma}
For comparison, our previously discussed  oracle Monte Carlo PIS estimator based
on $L$ independent draws $w_\ell\sim q_w$ has the standard $L^{-1/2}$ Monte Carlo rate: 
\[
\left|\hat\mu_{\mathrm{MC\text{-}PIS}}-\mu\right|
=
O_p(L^{-1/2}),
\]
provided $\operatorname{Var}_q(wm(w))<\infty$.
When this variance is not bounded,  the Monte Carlo convergence is
slower than $L^{-1/2}$, whereas the    $L^{-2}$ oracle quadrature rate \eqref{eq:quadrature_error} still remains valid.


\subsection{Consistency}
The preceding oracle analysis treats $m$ and $q_w$ as known.
We now account for their estimation error and establish
the consistency of PIS. Theorem~\ref{thm:A} provides general
consistency conditions for arbitrary regression and marginal
density estimators. Theorem~\ref{thm:unnormalized-consistency}
extends the result to unnormalized densities through
self-normalization. Theorem~\ref{thm:C} then verifies the
conditions for our proposed implementation using Gaussian
process regression and heavy-tailed density estimation.

\paragraph{General conditions for PIS consistency}
 We  first establish general conditions on these two components under which the resulting PIS estimator is consistent.

We first state our results for $p$ and $q$ normalized. The goal is to estimate the expectation $
\mu = \mathbb E_p[h(\theta)] = \int_0^\infty m(w)\,w\,q_w(w)\,dw.
$ Given proposal draws $\theta_{1:S}\sim q$, the general PIS framework estimates the conditional mean $m$ and marginal density $q_w$ by $\hat m$ and $\hat q_w$. We allow these two estimators to use the same training sample, with no restriction on their dependence. To account for numerical integration errors and truncation, we use the trapezoidal rule on a uniform grid of $L_S$ points over $[0,T_S]$:
\[
\hat\mu_{\mathrm{PIS}}
=
\operatorname{Quadrature}_{L_S}
\left[\hat m(w)w\hat q_w(w);[0,T_S]\right],
\]
where both the quadrature size $L_S$ and upper bound $T_S$
can depend on the sample size $S$.

\begin{theorem}[Consistency]
\label{thm:A}
Assume $\mathbb{E}_p\lvert h(\theta)\rvert < \infty$, the fitted $\hat{m}(w)$ and $\hat{q}_W(w)$ are Riemann integrable on compact sets, and $\hat{m}(w)$ is bounded in probability on compact sets. Assume
that, for every fixed $T < \infty$, as $S \to \infty$,
\begin{enumerate}[
    label=\textnormal{(\roman*)},
    ref=\textnormal{(\roman*)}
]
    \item \label{cond:regression-l1} \emph{The conditional mean estimate is $L_1$ consistent}
    $$
    \int_0^T
    \bigl\lvert \hat m(w)-m(w) \bigr\rvert
    q_w(w)\,\mathrm{d}w
    \xrightarrow{p} 0.
    $$
    
    \item \label{cond:density-l1} \emph{The marginal density  estimate is $L_1$ consistent}
    $$
    \int_0^T
    \bigl\lvert \hat q_w(w)-q_w(w) \bigr\rvert
    \,\mathrm{d}w
    \xrightarrow{p} 0.
    $$ 
\end{enumerate}
There exist deterministic sequences $T_S\uparrow\infty$ and $L_S\uparrow\infty$, with $L_S\ge S$, such that the PIS estimator is consistent in probability:
$$
\hat{\mu}_{\mathrm{PIS}} \xrightarrow{p} \mu.
$$
\end{theorem}

We provide a concrete quadrature scheme that truncate the integration at the largest observed importance weight.

\begin{lemma}[quadrature scheme]
\label{prop:sample-maximum}
Assume the conditions of Theorem~\ref{thm:A}. Suppose
$\Pr_q(W>w)=w^{-1/k_0}\ell(w)$ for all sufficiently large $w$,
where $\ell$ is slowly varying and $0<k_0<1$. For a fixed $c>0$, set
\[
M_S=\lfloor c\sqrt S\rfloor,\qquad
w_0=W_{(S-M_S)},\qquad
T_S=W_{(S)},\qquad
\pi=M_S/S.
\]
Fit $\hat q_w$ as in heavy tailed marginal model~\ref{def:tn-dpm-gpd} with this
threshold and tail fraction. Assume that, for some $\delta,\varepsilon>0$,
\[
\sup_{0\le w\le T_S}|\hat m(w)|=O_p(1),\qquad
\int_0^{w_0}w^{1+\delta}\hat q_{\mathrm{bulk}}(w)\,dw=O_p(1),
\]
and
\[
\hat\tau=O_p(w_0),\qquad
\Pr(0\le\hat k\le1-\varepsilon)\longrightarrow1.
\]
For the composite trapezoidal rule, there exists a deterministic
sequence $L_S\ge S$ such that
\[
\operatorname{Quadrature}_{L_S}
\left[\hat m(w)w\hat q_w(w);[0,T_S]\right]
\xrightarrow{p}\mu.
\]
\end{lemma}

We further extend this consistency result to the self-normalized PIS estimate:
\begin{theorem}[Consistency for unnormalized densities]
\label{thm:unnormalized-consistency}
Let $p$ and $q$ be unnormalized densities with unknown normalizing constants.
Assume that, on every fixed interval $[0,T]$, $\hat m$ and $\hat q_w$ satisfy the regression, density, boundedness, and Riemann-integrability conditions of Theorem~\ref{thm:A}. Then there exist deterministic sequences $T_S\uparrow\infty$ and
integers $L_S\geq S$, such that the self normalized PIS estimate is consistent, 
$$
\hat\mu_{\mathrm{PIS}}^{\mathrm{SN}}
=
\frac{
\operatorname{Quadrature}_{L_S}
\left[
\hat m(w)r\hat q_w(w);[0,T_S]
\right]
}{
\operatorname{Quadrature}_{L_S}
\left[
w\hat q_w(w);[0,T_S]
\right]
}
\xrightarrow{p}
\mu.$$
\end{theorem}

\paragraph{PIS consistency under our proposed regression and density estimation model.}

We now verify the consistency conditions for our recommended model-based implementation of Section~(\ref{sec:method}), using Gaussian process regression and a Pareto-tailed Dirichlet process mixture.  We take
$\hat m$ and $\hat q_w$ to be the corresponding posterior
mean estimates.

\begin{theorem}
\label{thm:C}
Assume $\mathbb E_p|h(\theta)|<\infty$ and that the distribution
of $W$ has support $[0,\infty)$. For a fixed $c>0$, set
$M=\lfloor c\sqrt S\rfloor$, $w_{0,S}=W_{(S-M)}$, and  $\pi=M/S$. Assume the following:
\begin{enumerate}[
    label=\textnormal{(\roman*)},
    ref=\textnormal{(\roman*)}
]

    \item \label{cond:C-gp}
    \emph{Conditional mean.}
    The observations satisfy
    $h_s=m(w_s)+\varepsilon_s$, where
    $\varepsilon_s\overset{\mathrm{iid}}{\sim}\mathrm \n(0,\sigma_0)$ independently of $w_{1:S}$, and $\sigma_0^2>0$ is known. We fit the observations to a Gaussian process regression  $m\sim\mathrm{GP}(m_0, \mathcal{K})$ and let $\hat m$ denote its
    posterior mean. Assume that $m_0$ is continuous, $\mathcal{K}$ is a continuous, nondegenerate, locally Lipschitz covariance kernel on
    $[0,\infty)^2$, and $m-m_0\in\mathcal H_K$.
    \item \label{cond:C-density}
    \emph{Marginal density.}
    Assume that $q_w$ is continuous and positive on every compact
    interval.
    At the threshold $w_0=w_{0,S}$, use the  Pareto-tailed Dirichlet-process mixture model \ref{def:tn-dpm-gpd} with
    $G\sim\operatorname{DP}(\alpha G_0)$ and
    $G_0=G_{0,\mu}\otimes G_{0,\sigma}$, where
    $$
    \begin{aligned}
    \mu
    &\sim \n(\mu_0,s_\mu)
    \mid 0\leq\mu\leq w_{0,S},\\
    \sigma^2
    &\sim \operatorname{InvGamma}(a_\sigma,b_\sigma)
    \mid 0<\sigma^2\leq\overline\sigma^2.
    \end{aligned}
    $$
    With hyperparameters $\mu_0\in\mathbb R$ and
    $\alpha,s_\mu,a_\sigma,b_\sigma,\overline\sigma>0$.
    Let $\hat q_{\mathrm{bulk}}$ denote the posterior mean
    based on $W_{(1)},\ldots,W_{(S-M)}$.

    \item \label{cond:C-tail}
    \emph{Marginal tail.}
    Assume that $q_w$ has an asymptotically Pareto tail: for some
    $0<k_0<1$,
    $$
    \frac{
    wq_w(w)
    }{
    \Pr(W>w)
    }
    \xrightarrow[]{w \xrightarrow{} \infty} 
    \frac{1}{k_0}.
    $$

\end{enumerate}

Then there exists a deterministic sequence $T_S\uparrow\infty$
and deterministic integers $L_S\geq S$ such that
$$
\hat\mu_{T_S,L_S}\xrightarrow{p}\mu.
$$

\end{theorem}

\section{Experiments}\label{sec:exp}
We compare PIS with self-normalized importance sampling
(SNIS) and Pareto-smoothed importance sampling
\citep[PSIS,][]{vehtari2024pareto} in simulated and real-data
experiments. We use the PSIS implementation in the
\texttt{loo} package \citep{loopackage}.

\subsection{Benchmark tasks}
We first test on the same benchmark tasks from \citet{vehtari2024pareto}, spanning a range of importance weight regimes, including the mean shift, scale mismatch, and distributional mismatch, all in a   $d=50$ dimensional space.

\begin{enumerate}
    \item $p(\theta)=\operatorname{MVN}(\mathbf{0}, I),\ q(\theta)=\operatorname{MVN}(0.4\cdot\mathbf{1}, 0.8I)$
    \item $p(\theta)=t_{20}(\mathbf{0}, I),\ q(\theta)=\operatorname{MVN}(0.4\cdot\mathbf{1}, 0.9I)$
    \item $p(\theta)=t_{20}(\mathbf{0}, I),\ q(\theta)=t_{21}(0.4\cdot\mathbf{1}, 0.8I)$
    \item $p(\theta)=t_{7}(\mathbf{0}, I),\ q(\theta)=t_{8}(0.4\cdot\mathbf{1}, 0.8I)$
    \item $p(\theta)=\operatorname{MVN}(\mathbf{0}, I),\ q(\theta)=t_{7}(0.4\cdot\mathbf{1}, 0.8I)$
\end{enumerate}

\begin{figure}[ht]
    \centering
    \includegraphics[width=\textwidth]{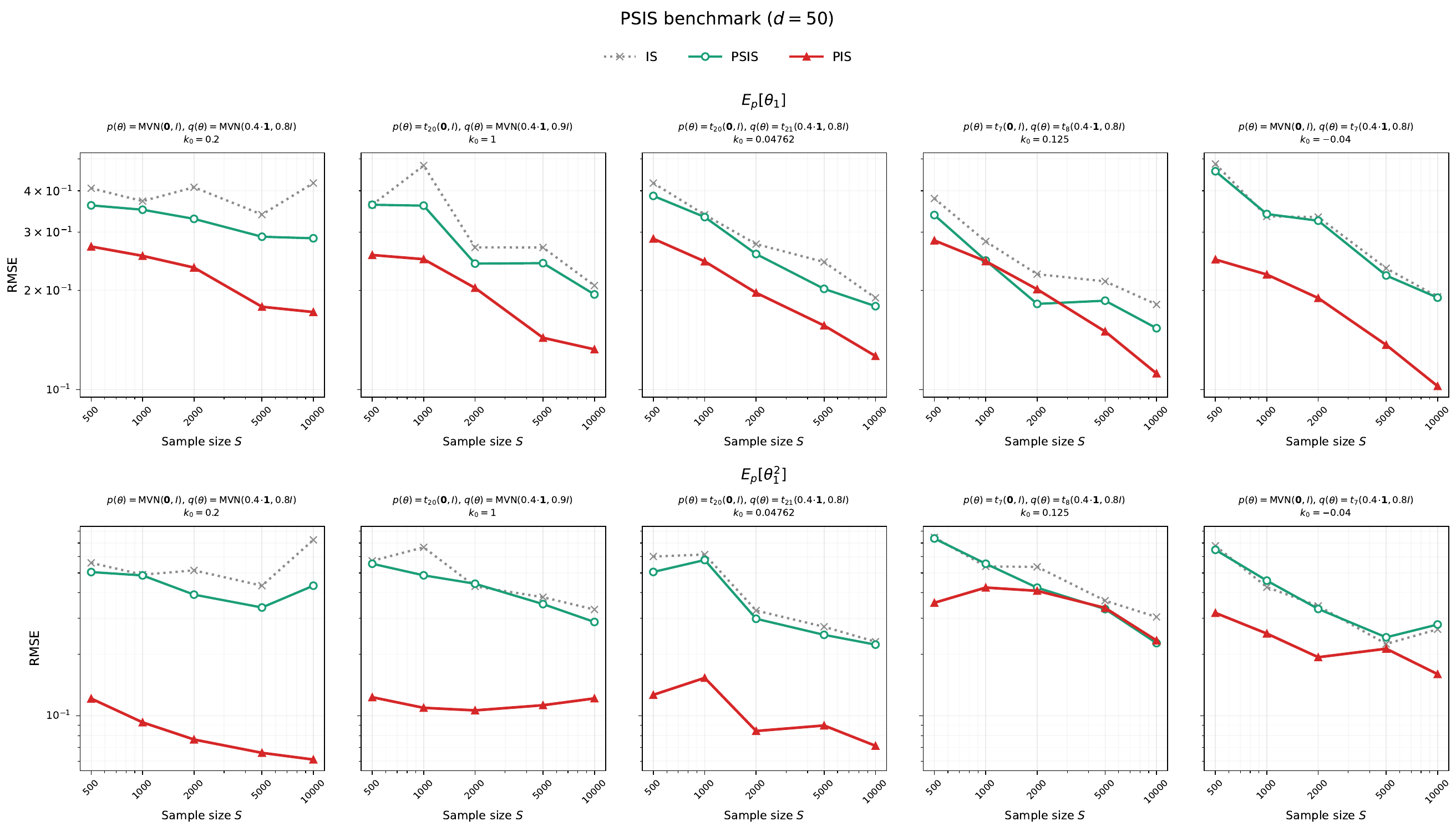}
    \caption{Comparison of PIS with regular importance sampling and PSIS on the benchmarking tasks. We report the root mean squared error (RMSE) of the estimated
    \(\mathbb E_p[\theta_1]\) (upper row), and   \(\mathbb E_p[\theta_1^2]\) (lower  row).}
    \label{fig:experiment2_linear}
\end{figure}

The upper row in Figure \ref{fig:experiment2_linear} presents the results for computing the expected linear functional \(\mathbb{E}_{p}[\theta_1]\). PIS generally achieves the lowest root mean squared error (RMSE), with substantial improvements over IS and PSIS in most configurations. 
The lower row presents the results for the second moment \(\mathbb{E}_{p}[\theta_1^2]\). The improvements from PIS are particularly pronounced in the first three configurations. 

\subsection{Larger gains in more challenging tasks}\label{exp:guassian}
To study how PIS can scale to increasing proposal-target-mismatch, and to higher dimensions, we consider a multivariate Gaussian target distribution $p(\theta)=\mbox{MVN}_d\!\left(\mathbf 0,(1-k_0)^{-1}I\right)$ and  proposal $q(\theta)=\mbox{MVN}_d(\mathbf 0,I)$. The corresponding importance ratio has a closed form density. 
\[
w(\theta)
=
\frac{p(\theta)}{q(\theta)}
=
(1-k_0)^{d/2}
\exp\!\left(\frac{k_0}{2}\|\theta\|^2\right).
\]
We do not supply this oracle marginal density nor the oracle   $k_0$ to any of the competing methods. Larger values of \(k_0\) correspond to a greater mismatch between the proposal and target distributions, and the variance of the weights is infinite when \(k_0\ge 1/2\).

We first estimate the target first moment $\mathbb E_p[\theta_1]=0$. 
We vary sample sizes $S\in\{500,1000,2000,5000,10000\}$
under two designs. First, we vary $k_0\in\{0.1,0.3,0.5,0.7,0.9\}$
while fixing \(d=10\). Second, we vary $d\in\{4,8,16,32,64\}$ separately for \(k_0\in\{0.25,0.55,0.75\}\). 

\begin{figure}[ht]
    \centering
    \includegraphics[width=\textwidth]{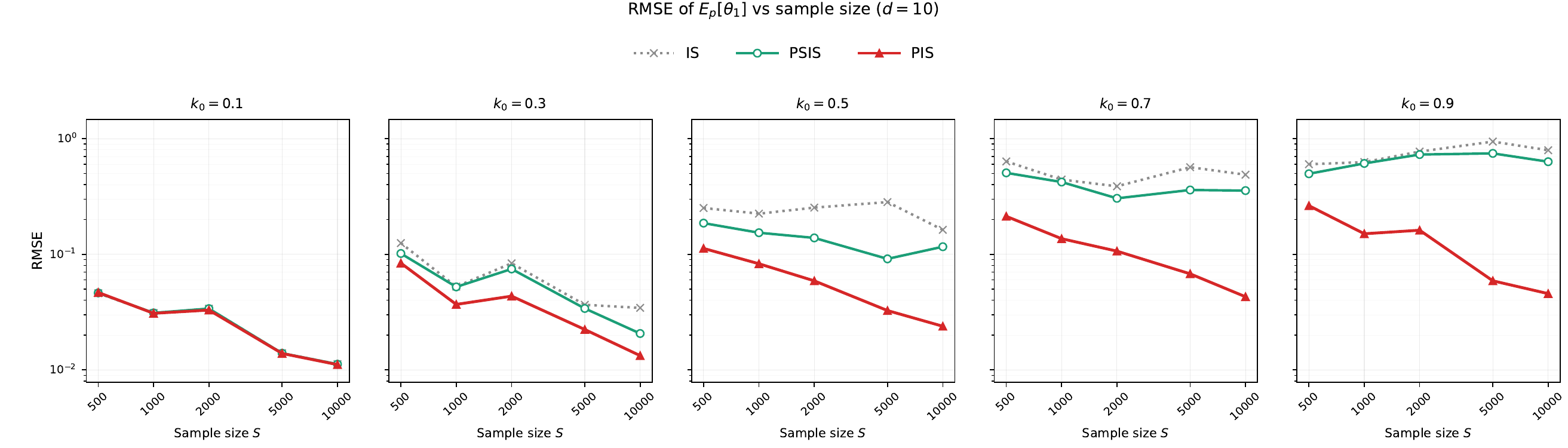}
    \caption{With both proposal and target being multivariate Gaussian, we compare PIS with self-normalized importance sampling  and Pareto smoothed importance sampling, measuring performance by RMSE of the estimated $\mathbb{E}_{p}[\theta_1]$. Larger $k$ indicates a harder problem.}
    \label{fig:experiment2_expectation}
\end{figure}

First, we vary the tail-heaviness.  Figure \ref{fig:experiment2_expectation} shows that in the easier, lighter-tailed cases with $k_0 < 0.5$, all methods perform similarly. But in heavier-tailed settings, the PIS estimator achieves noticeably lower RMSE than the competing methods. 

\begin{figure}[ht]
    \centering
    \includegraphics[width=\textwidth]{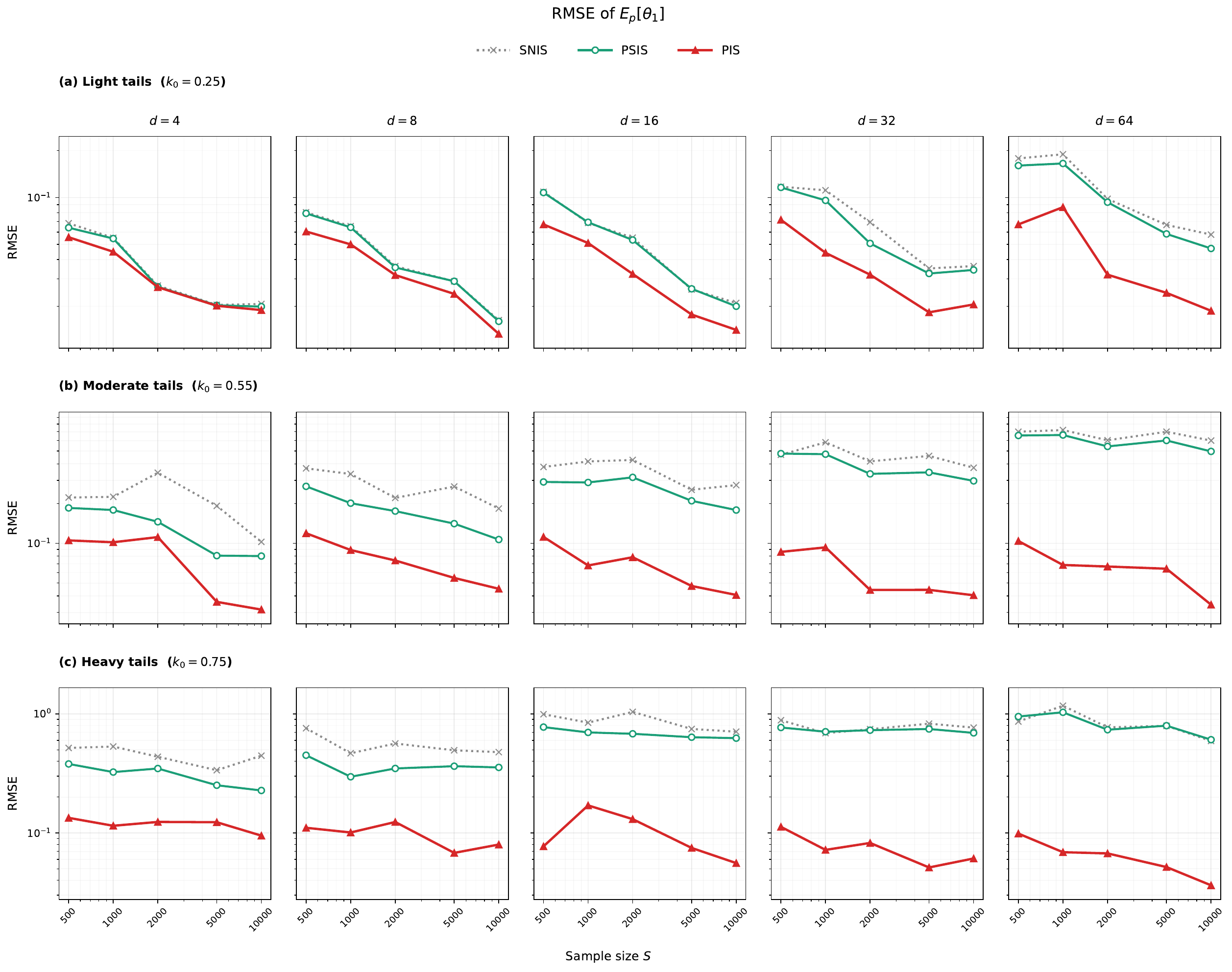}
    \caption{RMSE of the estimated
    $\mathbb{E}_{p}[\theta_1]$ under different tail heaviness and dimensions.
    PIS exhibits a larger gain in higher-dimensional tasks.}
    \label{fig:experiment2}
\end{figure}

We also vary the dimension of the configuration space.  Figure \ref{fig:experiment2} shows that the advantage of PIS becomes more pronounced as the dimension increases. Although the estimation problem becomes more challenging for all methods in higher dimensions, PIS maintains a lower RMSE and appears more stable than SNIS and PSIS, especially in higher dimensions. 


Next, we  use this Gaussian example to study the task of estimating a target tail probability:
\[
\mathbb E_p\!\left[\mathbbm 1\{\theta_1>2\}\right]
=
1-\Phi\!\left(2\sqrt{1-k_0}\right).
\]
Since $h(\theta)=\mathbbm 1\{\theta_1>2\}$ is binary, we use a Bernoulli-likelihood GP to estimate the conditional probability
$\Pr (\theta_1>2 |w)$.

\begin{figure}[ht]
    \centering
    \includegraphics[width=\textwidth]{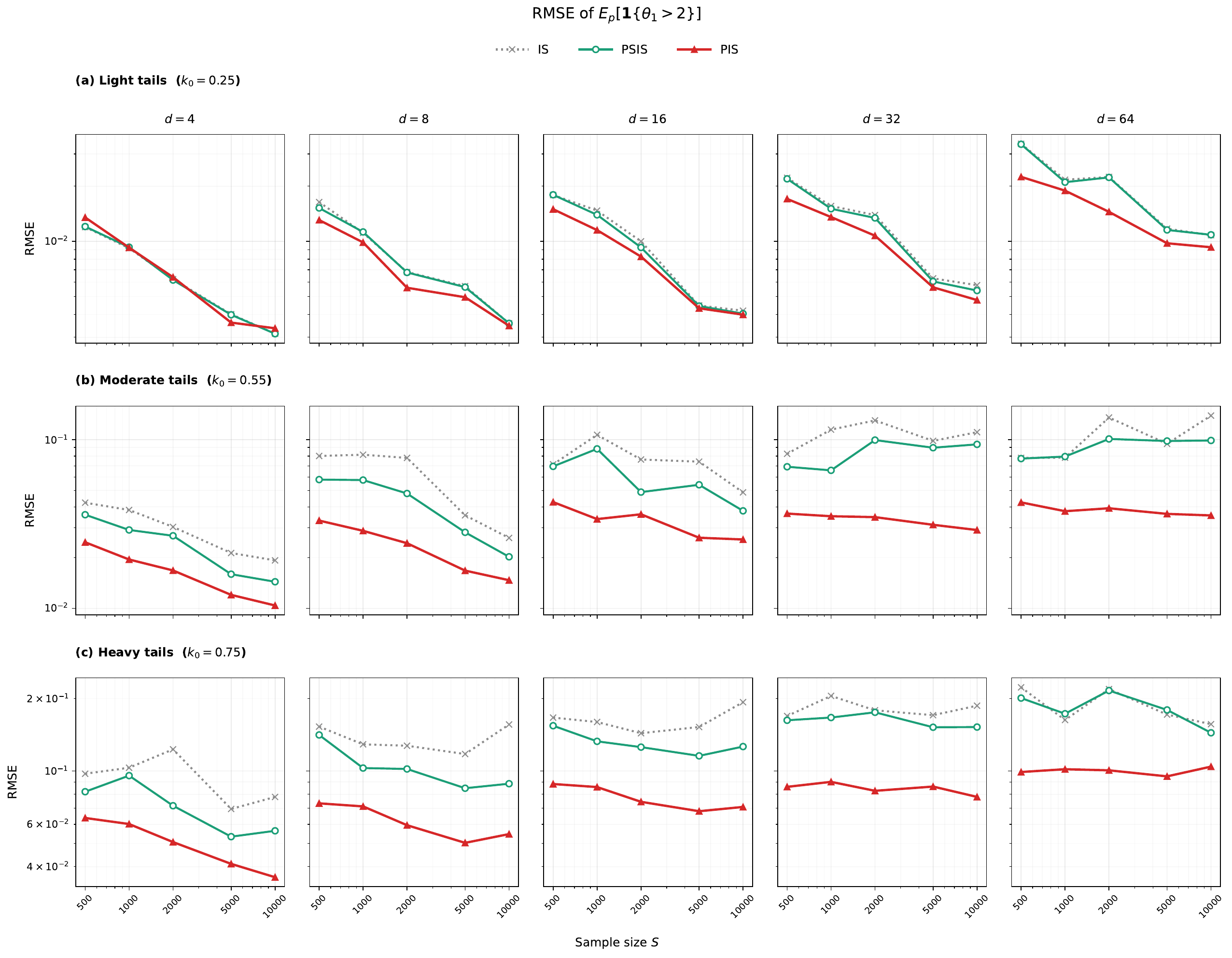}
    \caption{RMSE of the estimated
    $\mathbb{E}_{p}[1\{\theta_1>2\}]$
    under different tail regimes.}
    \label{fig:experiment4}
\end{figure}

Figure \ref{fig:experiment4} presents the RMSE of the tail probability estimate with varying $k_0$. In the light-tail case, PIS is consistently at least as good as PSIS and typically improves over both SNIS and PSIS, specifically in the high dimensional setting. In the moderate-tail regime ($k_0=0.55$), the importance weights have infinite variance, so the standard $S^{-1/2}$ Monte Carlo rate is no longer guaranteed. In the heavier-tail regime ($k_0=0.75$), the tail index exceeds the usual PSIS diagnostic threshold of $0.7$, a regime traditionally regarded as a practical no-go zone for importance sampling.   In both moderate- and heavy-tail regimes, all RMSE curves flatten, suggesting slower convergence for all methods. Nevertheless, PIS maintains substantially lower RMSE across the sample sizes and dimensions considered.

\subsection{Leave-one-out log predictive density, interval score, and coverage}\label{sec:LOO}
Leave-one-out (LOO) cross-validation is widely used to evaluate the predictive performance of Bayesian models \citep{vehtari2017practical}. Exact LOO evaluation requires refitting the model after removing each observation, which can be computationally expensive. Importance sampling avoids these repeated fits by reweighting draws from the full-data posterior \(p(\theta\mid y)\) toward the LOO posterior \(p(\theta\mid y_{-i})\), using weights
\[
w_i^{(s)}
\propto
\frac{p(\theta^{(s)}\mid y_{-i})}
     {p(\theta^{(s)}\mid y)}
\propto
\frac{1}{p(y_i\mid\theta^{(s)})}.
\]
However, these weights can be heavy-tailed when \(y_i\) is influential, making direct importance sampling unstable and potentially leading to infinite-variance estimators.
We use full-posterior draws to estimate  various LOO model evaluation metrics.
\begin{enumerate}
\item leave-one-out probability integral transformation \citep[PIT,][]{tesso2026loopitpredictivemodelchecking}, $\Pr (\tilde y_i < y_i \mid y_{_{-i}} ) $.
\item leave-one-out log predictive density $\log p(y_i | y_{-i})$; 
\item leave-one-out predictive coverage probability of an interval $\Pr (\tilde y_i\in [l, u] \mid  y_{-i})  $;
\item  leave-one-out interval score \citep{gneiting2007strictly} for the  $(\alpha/2,  1-\alpha/2)$ quantile interval;
\end{enumerate}
These evaluation metrics assess complementary aspects of predictive performance: local predictive accuracy, predictive coverage, and the joint calibration and sharpness of predictive intervals. We compare direct importance sampling, PSIS, and PIS, measuring their performance by squared error relative to the exact LOO. For the precise definitions, see Appendix~\ref{appendix:exp}.   

These metrics also represent interesting cases for the conditional mean model in PIS. For LOO log predictive density, $h(\theta)=p(y_i\mid\theta)\propto 1/w(\theta)$ is a known, deterministic function of the importance ratio $w$, so $m(w)$ is available exactly without regression. PIS therefore only needs to estimate the marginal weight density and perform one-dimensional quadrature. The other three metrics involve predictive event probabilities, which we estimate using binary indicators from posterior predictive draws and Bernoulli-likelihood GP regression. For the interval score, we first estimate the predictive CDF and invert it to obtain the required quantiles.

\begin{figure}[ht]
    \centering
    \includegraphics[width=\textwidth]    {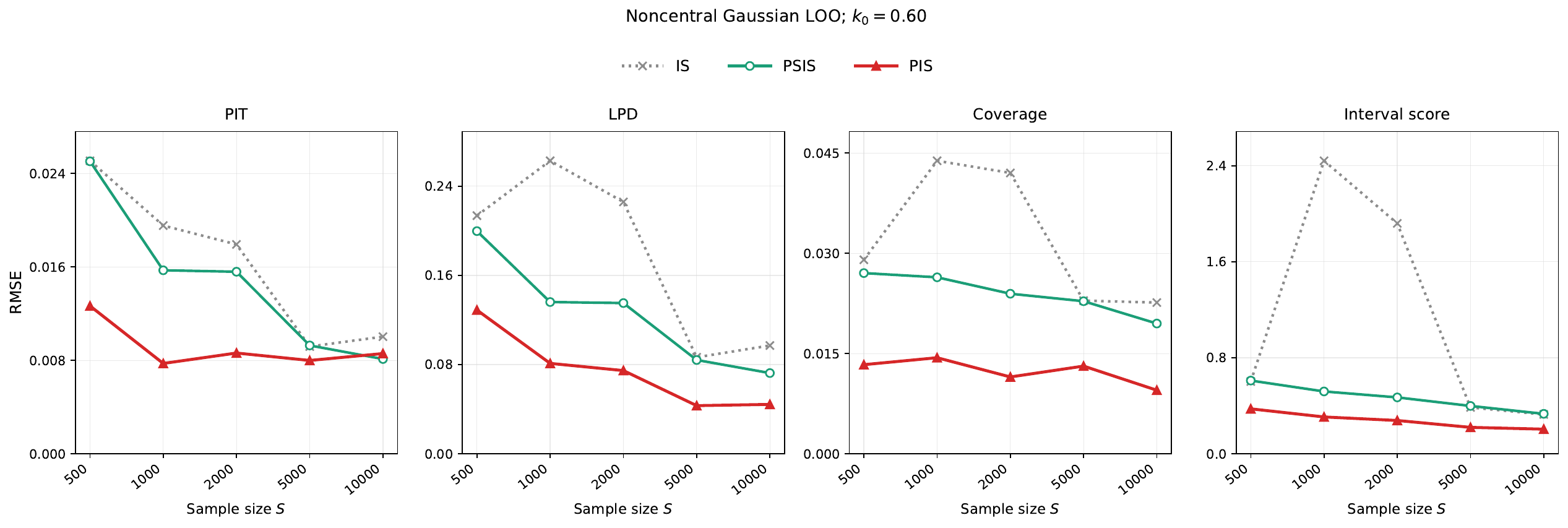}
    \caption{We use importance weighting to approximate leave-one-out metrics, including the probability integral transformation, log predictive density, coverage for a fixed interval $[-2.6, 2.6]$, and the interval score for $(0.05, 0.95)$ quantiles. In this Gaussian likelihood example, PIS generally outperforms PSIS and IS.}
    \label{fig:experiment3_all_panels}
\end{figure}

For an analytically tractable LOO benchmark, consider a Gaussian model with likelihood
$y_i\mid\theta\sim\n(\theta,1)$. We fix one held-out observation $y_i=2$.  Suppose that the true LOO posterior, conditional on the remaining observations, is
$
p(\theta)\coloneqq p(\theta\mid y_{-i})=\n(0,\sqrt{1.5}).
$
The available 
full-data posterior is 
$
q(\theta) \coloneqq  p(\theta\mid y)
=\n\left(1.2,\sqrt{0.6}\right).
$
The importance weight is the inverse-likelihood 
$
w_i(\theta)
\coloneqq \frac{1}{p(y_i\mid\theta)}
=\sqrt{2\pi}\exp\left(\frac{(\theta-2)^2}{2}\right),
$
which has a Pareto tail index $k=0.6$  under the full-data posterior and hence infinite variance.
 Figure \ref{fig:experiment3_all_panels} shows that PIS achieves the lowest RMSE
for all four predictive metrics. The improvement is particularly pronounced
for log predictive densities and interval scores, where ordinary importance sampling is highly unstable. 

\subsection{Calibrating approximate inference}
When approximate inference is used for Bayesian posterior computation, it is often desirable to correct the approximation error \citep[e.g.,][]{agrawal2020advances, goplerud2022fast}. Importance sampling provides a natural calibration by reweighting draws from an approximate posterior \(q(\theta)\) toward the target  \(p(\theta |  y)\). However, this correction can be unstable, especially when \(q\) is lighter-tailed, as is common for variational approximation.

This application provides a realistic but challenging benchmark for proposal and target distributions. To evaluate PIS in this setting, we run variational inference on a collection of Stan models from PosteriorDB \citep{magnusson2024posteriordb}. For each model, we fit a mean-field variational approximation, use its draws as proposal samples, and compare the estimates of the target posterior expectations  obtained from direct importance sampling, PSIS, and PIS, where their performance is measured by the  squared error relative to a long-run MCMC.

\begin{figure}[h]
    \centering
    \includegraphics[
        width=\textwidth
    ]{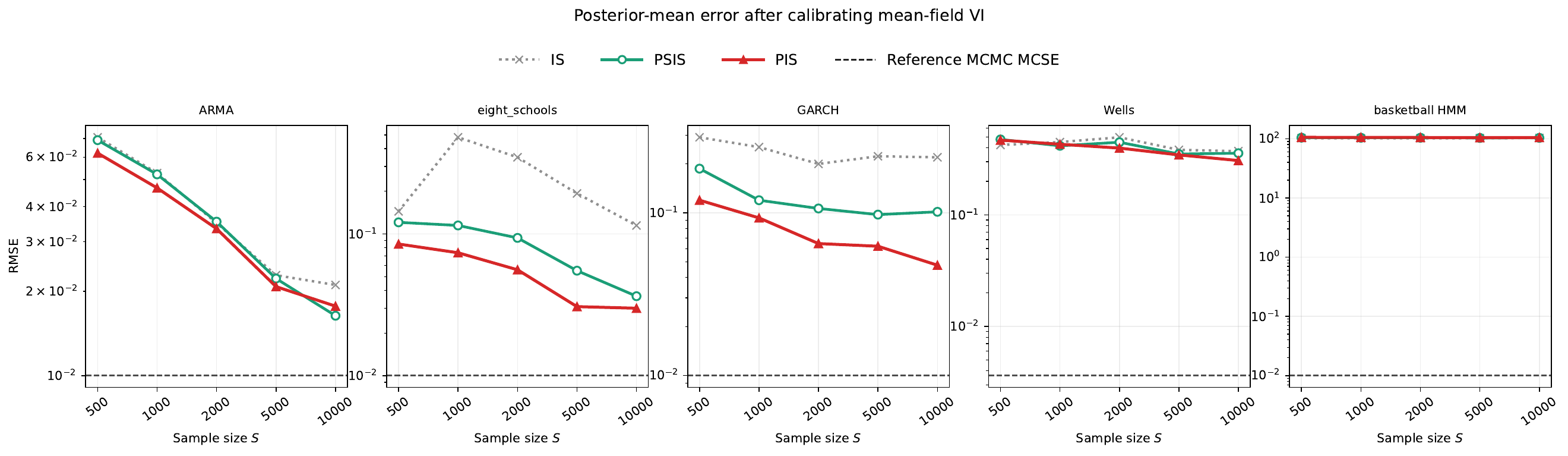}
    \caption{
        Posterior-mean estimation error after calibrating mean-field variational approximations for five PosteriorDB models.
    }
    \label{fig:posteriordb-calibration}
\end{figure}

Figure~\ref{fig:posteriordb-calibration} compares IS, PSIS, and PIS across PosteriorDB models with varying degrees of
target-proposal mismatch. For the \texttt{ARMA} model, the importance weights are light-tailed, and PIS offers only modest gains.
For \texttt{eight\_schools} and \texttt{GARCH}, PIS substantially reduces estimation error, especially as the sample size increases. For \texttt{Wells}, the extremely heavy tails limit the rate of improvement. Finally, in the \texttt{basketball HMM}, mean-field VI fails to capture the typical set,  leaving all three methods with persistent bias. This failure mode illustrates a limitation
of PIS: severe target-proposal mismatch cannot necessarily be corrected by poststratification.


\subsection{Uncertainty quantification}
Reliable importance sampling requires not only accurate point estimates but also uncertainty intervals with appropriate coverage. PIS directly reports the Bayesian posterior distribution of the mean estimate. We study the same Gaussian proposal and target in Sec~\ref{exp:guassian}. We use IS, PSIS, and PIS to estimate the expectation $\mu=\mathbb E_p[\theta_1]$ and its confidence interval. We repeat the simulations and evaluate the empirical coverage of the reported confidence intervals. 

\begin{figure}[h]
    \centering
    \includegraphics[
        width=0.75\textwidth
    ]{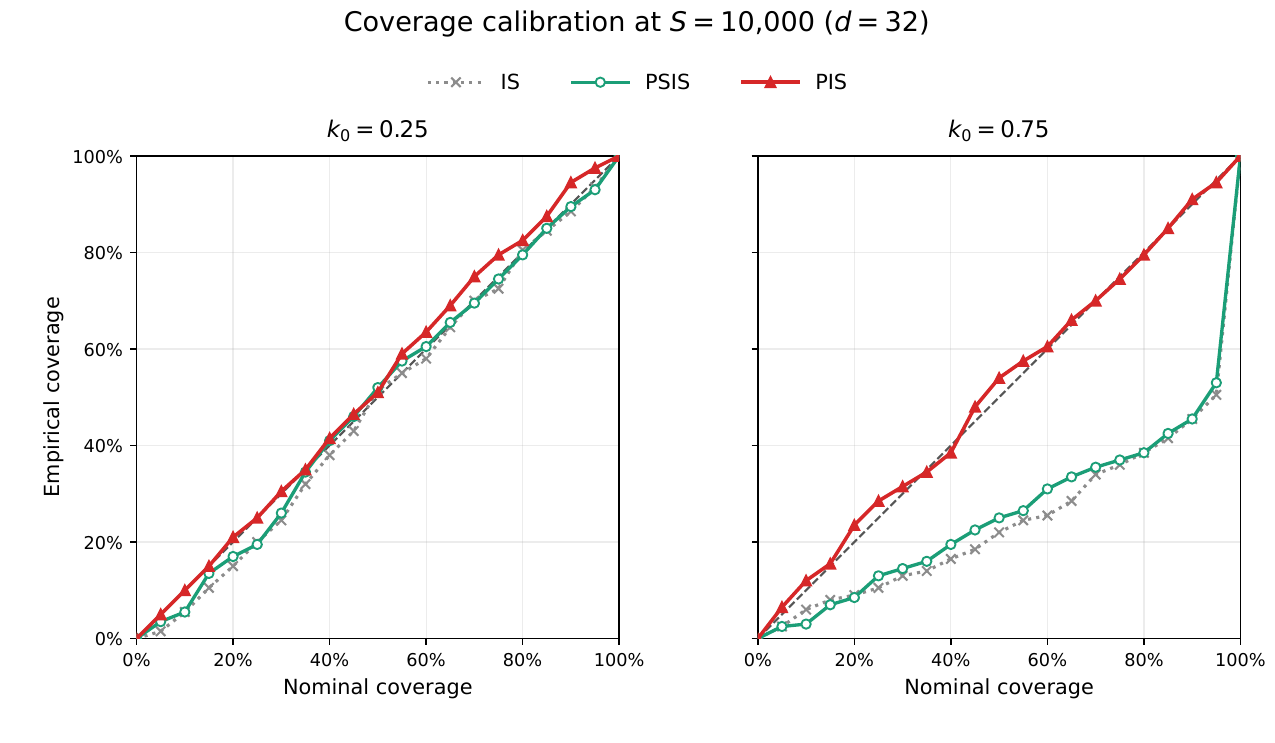}
    \caption{
        Empirical versus nominal coverage for intervals from three methods. Perfect calibration is the diagonal line.}
    \label{fig:coverage-calibration-k025-k075}
\end{figure}

As a caveat, PIS reports Bayesian credible intervals,
which are not guaranteed to achieve nominal frequentist coverage. Nevertheless, Figure \ref{fig:coverage-calibration-k025-k075} evaluates this frequentist calibration  
at \(S=10{,}000\) and \(d=32\). When \(k_0=0.25\), all three methods are approximately calibrated. When \(k_0=0.75\), the confidence intervals from SNIS and PSIS are severely overconfident, whereas PIS remains close to the ideal calibration. 

\begin{figure}[h]
    \centering
    \includegraphics[
        width=0.75\textwidth
    ]{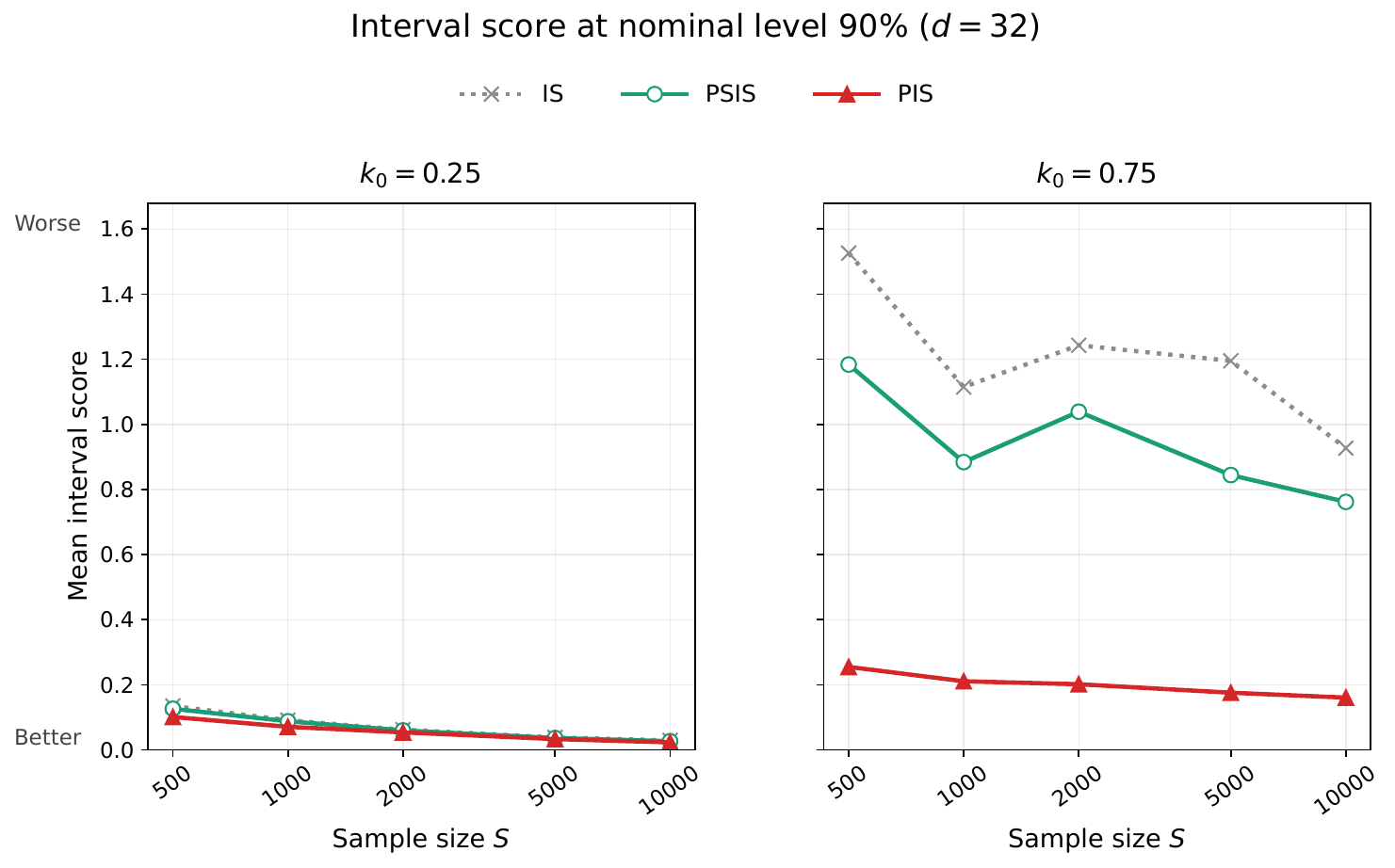}
    \caption{
        Mean interval score for nominal $90\%$ intervals as a function
        of sample size at $d=32$. 
    }
    \label{fig:interval-score-k025-k075}
\end{figure}

Figure \ref{fig:interval-score-k025-k075} reports the mean interval score for the reported 90\% confidence interval, which penalizes both interval width and
failure to cover the true expectation. Lower scores indicate a better overall trade-off between confidence interval sharpness and coverage. At \(k_0=0.25\), all methods improve with sample size and perform similarly, with a small advantage for PIS. At \(k_0=0.75\), PIS exhibits a significant advantage, while PSIS improves upon SNIS but remains considerably worse than PIS.  Together, these results suggest that PIS improves not only point estimation but also uncertainty quantification, particularly in heavy-tailed settings where conventional importance sampling standard error becomes unreliable.

\section{Discussion}

\subsection{Heuristics from survey sampling}
Importance weighting \eqref{eq:IS} has a direct analog in survey sampling. In an observational study where the survey subjects differ from the population of interest, two orthogonal approaches exist to adjust for the population mismatch. The first approach, inverse-probability weighting, reweighs sampled outcomes using inverse inclusion probabilities, which is effectively importance sampling from the survey population to the target.

An alternative strategy in survey sampling is multilevel regression and poststratification \citep[MRP,][]{little1993poststratification, gelman1997poststratification}. MRP groups the survey observations into   \emph{strata} or \emph{cells} defined by  demographic  features (such as age, gender,  state, education), and then sums over estimated strata-level outcomes.  Concretely, let $y$ denote an individual survey outcome such as voting preference, and $x\in \mathcal{X}$ denote demographic variables. MRP uses a similar law of total expectation identity:
\begin{equation}\label{eq:MRP}
\E_p[y]
=
\E_p [\E_p[y \mid x]]
=
\sum_{x_i \in \mathcal{X}} \left(\Pr(x=x_i)\E_p[y\mid x] \right)  
\end{equation}
As the name suggests, MRP performs two steps: (i) fit a model-based, typically multilevel regression to estimate \(\E_p[y\mid x]\) to the survey observations, and (ii) average over $\Pr(x)$, the population distribution of  demographic $x$, typically known from a census table. It is not surprising that MRP has become a workhorse in survey sampling, with applications  such as in political science \citep{lax2009should}, opinion polling \citep{ghitza2013deep, wang2015forecasting}, and epidemiology \citep{gelman2020bayesian, covello2021routine}. It is perhaps more surprising that this conditional-modeling framework \eqref{eq:MRP} was not fully developed for the general importance sampling problem---Our present paper takes this step.

Despite the shared heuristic, our paper is not a direct application of MRP to statistical computing. First, the MRP poststratification variables $x$ are externally chosen, often discrete, and their target population distribution is typically known from a census table. PIS  instead conditions on the  scalar importance ratio $w$ while its target distribution \(p_w(w)\) needs to be estimated. Second, the identity of MRP \eqref{eq:MRP} relies on a non-testable ignorability assumption: conditional on the poststratification variables, sample inclusion is \emph{assumed} to be independent of the outcome. In contrast, the analogous conditional invariance in PIS is by construction (Thm.~\ref{prop:conditional_equivalence}).
Third, in PIS the weight $w$ is a deterministic function of \(\theta\), so that mathematically the PIS identity \eqref{eq_strata}  requires a  geometric measure theoretic formulation. Though conceptually more involved than MRP, it enables score-based marginal density estimation. 
Lastly, other than the empirical evidence \citep[e.g.,][]{downes2018multilevel}, 
formal theory comparing MRP with weighting remains limited.  Our theoretical analysis quantifies the gains from conditional modeling over direct weighting in importance sampling, and may in turn motivate analogous theory in survey sampling.


\subsection{Relation to other importance sampling methods}
\paragraph{Regularized importance sampling.}
There is  an existing line of research on stabilizing importance sampling by regularizing the realized weights, and many of the proposed methods can now correspond to particular model choices in the PIS framework. 
Examples include truncated importance sampling \citep{ionides2008truncated}
and Pareto-smoothed importance sampling
\citep[PSIS;][]{vehtari2024pareto}. PSIS stabilizes importance sampling by fitting a generalized Pareto distribution to the tail of the observed importance ratios and replacing the largest few raw weights with expected order statistics.
This tail modeling in PSIS inspires the same generalized Pareto transformation in our density architecture.  Rather than smoothing only the largest weights, PIS estimates a full generative model of $w$, including both the bulk and tail. In light of the general PIS framework, PSIS corresponds to using an interpolating estimate of the conditional mean \(m(w_s)\approx h(\theta_s)\), while modeling the marginal of \(w\) by an empirical distribution in the bulk and a generalized Pareto approximation in the tail.
Given that PSIS is currently used as the state-of-the-art weighting tool in the current Bayesian workflow \citep{gelman2020bayesianflow}, our empirical comparisons treat PSIS as the primary baseline, and we generally find that our model-based PIS improves upon PSIS in both accuracy and uncertainty assessment.

\paragraph{Adaptive proposal construction.}
The term \emph{post}-stratified in our method emphasizes that the sampling distribution \(q\) is taken as given, in the same sense that \emph{post}-stratification in survey sampling makes inference after a survey has already been collected rather than drawing a new survey. Likewise, PIS aims to improve the reweighting task when the proposal \(q\) is fixed. This setting often arises in Bayesian computation: the proposal may be an existing posterior sample reused for cross-validation, or a variational sample in an approximate inference correction. 
However, in other settings when one can instead design the proposal and generate additional samples therein, an orthogonal line of research seeks to construct or adapt \(q\) so that the resulting weighting task can be easier. Examples to design a better $q$ include mixture proposals \citep{owen2000safe, silva2024robust}, adaptive and population Monte Carlo \citep{cappe2008adaptive,cornuet2012adaptive,bugallo2017adaptive,elvira2019generalized}, 
target-aware split proposal \citep{rainforth2020target}, 
umbrella sampling \citep{dinner2020stratification},
variational inference with chi squared divergence \citep{dieng2017chi} or importance weighted divergence \citep{domke2018importance}, and affine transformation to match moments \citep{paananen2021implicitly}. 
Combining proposal adaptation with PIS is a natural direction
for future work: adaptation improves the proposal sample, while PIS improves inference given that sample.

\subsection{Why condition on the importance ratio?}
PIS conditions on the importance ratio $w=r(\theta)$. In principle, we can extend this framework to allow any conditioning variables $T=T(\theta)\in\mathbb R^k$, $k\leq d$ in the tower identity
\[
\mathbb E_p[h(\theta)]
=
\int_{\mathbb R^k}
\mathbb E_p[h(\theta)\mid T=t]\,p_T(t)\, dt.
\]
Choosing $T(\theta)=r(\theta)$ ensures conditional invariance,
\[
\mathbb E_p[h(\theta)\mid w]
=
\mathbb E_q[h(\theta)\mid w],
\]
for every integrable $h$, so the conditional expectation under $p$ can be learned from proposal draws without additional weighting.

Theorem~\ref{prop:conditional_equivalence} indicates that any conditioning variable $T$ with this invariance property must determine $w$: there exists a measurable function $\alpha$ such that $w=\alpha(T)$ almost surely. Thus, $w$ is the coarsest conditioning variable with this property up to a transformation. 




\subsection{Limitations and future work} 
As an approximation method, PIS has limitations and failure modes. First, our current PIS formulation assumes that the target and proposal densities \(p\) and \(q\) are smooth and have full support on $\R^d$. While it covers many continuous Bayesian models that we are interested in, it excludes settings with discrete parameters, bounded or truncated supports, or piecewise-constant approximations. When the support is discrete or truncated, the same PIS identity still holds, but the marginal density of $w$ would need to be replaced by a discrete or mixed discrete-continuous estimate.  We avoid this generality in the present paper for brevity. 

Second, the present paper focuses on expectations. This task is distinct from \emph{importance resampling} \citep{rubin1987sir}, which aims to \emph{generate} approximate target draws from the available proposal draws. Importance resampling carries the same instability as importance weighting: a few realized draws with large weights can dominate the resampled population. Moreover, the resulting sample has fixed support on the original proposal draws. It is a promising direction to extend PIS to the generative sampling task, which would require modeling the conditional law of \(\theta\mid w\), and is left for future work.

Third, as in much of the importance sampling literature, we have assumed IID  proposal draws. If instead they were obtained by MCMC, autocorrelation can bias both the regression estimate \(m(w)\) and the marginal-density estimate \(q_w(w)\). Before PIS is extended to dependent sampling in future work, thinning  is recommended before using PIS when autocorrelation is severe.



Lastly, the one-dimensionality of PIS does not fully eliminate the curse of dimensionality. Although PIS reduces the task to estimating the one-dimensional functions \(m(w)\) and \(p_w(w)\), the quality of these estimates still depends on the proposal sample generated in the original \(d\)-dimensional space. When the dimension is high, or when the proposal severely mismatches the target, this dependence can appear in three ways: (a) The marginal $w$  distribution may become increasingly heavy-tailed, making the marginal density harder to estimate; (b) When the heavy tail is represented by only a few observed proposal draws, the regression estimate locally relies on model extrapolation and prior regularization, and may exhibit a large uncertainty; (c) When $w$ is heavy-tailed, the integrand \(m(w)p_w(w)\) can decay slowly, so the numerical integration may require a wider and denser  quadrature grid in the tail to control the truncation error.

Looking forward, the development of PIS represents an instance of bringing the Bayesian modeling workflow \citep{gelman2020bayesianflow} into Monte Carlo computation.  
This echoes a longstanding critique of conventional Monte Carlo integration \citep{ohagan1987monte}: despite its universal applicability, its averaging rule does not adapt to the observed integrand or exploit its underlying structure. 
More broadly, this reflects a contrast between the two  cultures in statistics  \citep{gelman2021reflections}: simple-form-for-robustness  versus model-as-much-as-you-can. 
Conventional IS is not entirely model-free: within the PIS framework, it implicitly uses an empirical weight distribution and an unsmoothed conditional mean. PIS makes these modeling choices explicit, allowing them to be regularized, checked, and iteratively improved through the modeling workflow. We believe this modeling perspective can
also benefit other computational methods traditionally treated as fixed algorithms.

\small
\bibliographystyle{apalike}
\bibliography{main}	

@article{gelman1997poststratification,
  title={Poststratification into many categories using hierarchical logistic regression},
  author={Gelman, Andrew and   Thomas  Little},
  journal={Survey Methodology},
  volume={23},
  pages={127},
  year={1997}
}

@article{ghosal1999posterior,
  author  = {Ghosal, Subhashis and Ghosh, Jayanta K. and
             Ramamoorthi, R. V.},
  title   = {Posterior Consistency of Dirichlet Mixtures in
             Density Estimation},
  journal = {The Annals of Statistics},
  year    = {1999},
  volume  = {27},
  number  = {1},
  pages   = {143--158},
  doi     = {10.1214/aos/1018031105}
}

@book{atkinson1989introduction,
  author    = {Atkinson, Kendall E.},
  title     = {An Introduction to Numerical Analysis},
  edition   = {2},
  publisher = {John Wiley \& Sons},
  year      = {1989}
}

@book{kallenberg2002foundations,
  author    = {Kallenberg, Olav},
  title     = {Foundations of Modern Probability},
  edition   = {2},
  year      = {2002},
  publisher = {Springer},
  address   = {New York},
  doi       = {10.1007/978-1-4757-4015-8}
}

@article{tokdar2006posterior,
    author  = {Tokdar, Surya T.},
    title   = {Posterior Consistency of Dirichlet Location-Scale Mixture of Normals in Density Estimation and Regression},
    journal = {Sankhyā: The Indian Journal of Statistics},
    year    = {2006},
    volume  = {68},
    number  = {1},
    pages   = {90--110},
    url     = {https://www.jstor.org/stable/25053478}
}

@article{koepernik2021consistency,
  author  = {Koepernik, Peter and Pfaff, Florian},
  title   = {Consistency of Gaussian Process Regression in Metric Spaces},
  journal = {Journal of Machine Learning Research},
  year    = {2021},
  volume  = {22},
  number  = {244},
  pages   = {1--27},
  url     = {https://jmlr.org/papers/v22/21-0853.html}
}

@article{ghitza2013deep,
  title={Deep interactions with {MRP}: Election turnout and voting patterns among small electoral subgroups},
  author={Ghitza, Yair and Gelman, Andrew},
  journal={American Journal of Political Science},
  volume={57},
  number={3},
  pages={762--776},
  year={2013}
}

@article{wang2015forecasting,
  title={Forecasting elections with non-representative polls},
  author={Wang, Wei and Rothschild, David and Goel, Sharad and Gelman, Andrew},
  journal={International Journal of Forecasting},
  volume={31},
  number={3},
  pages={980--991},
  year={2015}
}

@article{lax2009should,
  title={How should we estimate public opinion in the states?},
  author={Lax, Jeffrey R and Phillips, Justin H},
  journal={American Journal of Political Science},
  volume={53},
  number={1},
  pages={107--121},
  year={2009},
  publisher={Wiley Online Library}
}

@article{downes2018multilevel,
  title={Multilevel regression and poststratification: a modeling approach to estimating population quantities from highly selected survey samples},
  author={Downes, Marnie and Gurrin, Lyle C and English, Dallas R and Pirkis, Jane and Currier, Dianne and Spittal, Matthew J and Carlin, John B},
  journal={American Journal of Epidemiology},
  volume={187},
  number={8},
  pages={1780--1790},
  year={2018},
  publisher={Oxford University Press}
}

@article{gelman2020bayesian,
  title={Bayesian analysis of tests with unknown specificity and sensitivity},
  author={Gelman, Andrew and Carpenter, Bob},
  journal={Journal of the Royal Statistical Society Series C},
  volume={69},
  number={5},
  pages={1269--1283},
  year={2020},
  publisher={Oxford University Press}
}

@book{evans2015measure,
  title={Measure Theory and Fine Properties of Functions},
  author={Evans, Lawrence C. and Gariepy, Ronald F.},
  year={2015},
  publisher={CRC Press}

}

@article{greengard2025equispaced,
  title={Equispaced {F}ourier representations for efficient {G}aussian process regression from a billion data points},
  author={Greengard, Philip and Rachh, Manas and Barnett, Alex H},
  journal={SIAM/ASA Journal on Uncertainty Quantification},
  volume={13},
  number={1},
  pages={63--89},
  year={2025},
  publisher={SIAM}
}

@article{vehtari2024pareto,
  title={Pareto smoothed importance sampling},
  author={Vehtari, Aki and Simpson, Daniel and Gelman, Andrew and Yao, Yuling and Gabry, Jonah},
  journal={Journal of Machine Learning Research},
  volume={25},
  number={72},
  pages={1--58},
  year={2024}
}

@article{vehtari2017practical,
  title={Practical {B}ayesian model evaluation using leave-one-out cross-validation and {WAIC}},
  author={Vehtari, Aki and Gelman, Andrew and Gabry, Jonah},
  journal={Statistics and Computing},
  volume={27},
  number={5},
  pages={1413--1432},
  year={2017},
  publisher={Springer}
}

@article{elvira2022rethinking,
  title={Rethinking the effective sample size},
  author={Elvira, V{\'\i}ctor and Martino, Luca and Robert, Christian P},
  journal={International Statistical Review},
  volume={90},
  number={3},
  pages={525--550},
  year={2022},
  publisher={Wiley Online Library}
}

@article{hooten2021making,
  title={Making recursive {B}ayesian inference accessible},
  author={Hooten, Mevin B and Johnson, Devin S and Brost, Brian M},
  journal={The American Statistician},
  volume={75},
  number={2},
  pages={185--194},
  year={2021},
  publisher={Taylor \& Francis}
}

@incollection{doucet2001introduction,
  title={An introduction to sequential {M}onte {C}arlo methods},
  author={Doucet, Arnaud and De Freitas, Nando and Gordon, Neil},
  booktitle={Sequential Monte Carlo Methods in Practice},
  pages={3--14},
  year={2001},
  publisher={Springer}
}

@article{chopin2002sequential,
  title={A sequential particle filter method for static models},
  author={Chopin, Nicolas},
  journal={Biometrika},
  volume={89},
  number={3},
  pages={539--552},
  year={2002},
  publisher={Oxford University Press}
}

@article{little1993poststratification,
  title   = {Post-Stratification: A Modeler's Perspective},
  author  = {Little, Roderick J. A.},
  journal = {Journal of the American Statistical Association},
  volume  = {88},
  number  = {423},
  pages   = {1001--1012},
  year    = {1993},
  doi     = {10.1080/01621459.1993.10476368}
}

@article{goplerud2022fast,
  title={Fast and Accurate Estimation of Non-Nested {B}inomial Hierarchical Models Using Variational Inference},
  author={Goplerud, Max},
  journal={Bayesian Analysis},
  volume={17},
  number={2},
  pages={623--650},
  year={2022}
  }

@article{gelfand1992model,
  title={Model determination using predictive distributions with implementation via sampling-based methods},
  author={Gelfand, Alan E and Dey, Dipak K and Chang, Hong},
  journal={Bayesian Statistics},
  volume={4},
  pages={147--167},
  year={1992}
}

@article{peruggia1997variability,
  title={On the variability of case-deletion importance sampling weights in the {B}ayesian linear model},
  author={Peruggia, Mario},
  journal={Journal of the American Statistical Association},
  volume={92},
  number={437},
  pages={199--207},
  year={1997},
  doi={10.1080/01621459.1997.10473617}
}

@article{paananen2021implicitly,
  title={Implicitly adaptive importance sampling},
  author={Paananen, Topi and Piironen, Juho and B{\"u}rkner, Paul-Christian and Vehtari, Aki},
  journal={Statistics and Computing},
  volume={31},
  number={2},
  pages={16},
  year={2021},
  doi={10.1007/s11222-020-09982-2}
}

@inproceedings{magnusson2019bayesian,
  title={Bayesian leave-one-out cross-validation for large data},
  author={Magnusson, M{\aa}ns and Andersen, Michael Riis and Jonasson, Johan and Vehtari, Aki},
  booktitle={International Conference on Machine Learning},
  volume={97},
  pages={4244--4253},
  year={2019}
}

@article{burkner2020approximate,
  title={Approximate leave-future-out cross-validation for {B}ayesian time series models},
  author={B{\"u}rkner, Paul-Christian and Gabry, Jonah and Vehtari, Aki},
  journal={Journal of Statistical Computation and Simulation},
  volume={90},
  number={14},
  pages={2499--2523},
  year={2020},
  doi={10.1080/00949655.2020.1783262}
}

@article{smith1992bayesian,
  title={Bayesian statistics without tears: {A} sampling-resampling perspective},
  author={Smith, Adrian F M and Gelfand, Alan E},
  journal={The American Statistician},
  volume={46},
  number={2},
  pages={84--88},
  year={1992}
}

@article{kallioinen2024detecting,
  title={Detecting and diagnosing prior and likelihood sensitivity with power-scaling},
  author={Kallioinen, Noa and Paananen, Topi and B{\"u}rkner, Paul-Christian and Vehtari, Aki},
  journal={Statistics and Computing},
  volume={34},
  number={57},
  year={2024},
  doi={10.1007/s11222-023-10366-5}
}

@inproceedings{burda2016importance,
  title={Importance Weighted Autoencoders},
  author={Burda, Yuri and Grosse, Roger and Salakhutdinov, Ruslan},
  booktitle={International Conference on Learning Representations},
  year={2016}
}

@inproceedings{domke2018importance,
  title={Importance weighting and variational inference},
  author={Domke, Justin and Sheldon, Daniel R},
  booktitle={Advances in Neural Information Processing Systems},
  volume={31},
  year={2018}
}

@inproceedings{yao2018yes,
  title={Yes, but did it work?: {E}valuating variational inference},
  author={Yao, Yuling and Vehtari, Aki and Simpson, Daniel and Gelman, Andrew},
  booktitle={International Conference on Machine Learning},
  pages={5581--5590},
  year={2018},
  organization={PMLR}
}

@article{magnusson2024posteriordb,
  title={posteriordb: Testing, benchmarking and developing {B}ayesian inference algorithms},
  author={Magnusson, M{\aa}ns and Torgander, Jakob and B{\"u}rkner, Paul-Christian and Zhang, Lu and Carpenter, Bob and Vehtari, Aki},
  journal={arXiv preprint arXiv:2407.04967},
  year={2024}
}

@article{chatterjee2018sample,
  title   = {The Sample Size Required in Importance Sampling},
  author  = {Chatterjee, Sourav and Diaconis, Persi},
  journal = {Annals of Applied Probability},
  year    = {2018},
  volume  = {28},
  number  = {2},
  pages   = {1099--1135},
  doi     = {10.1214/17-AAP1326}
}

@inproceedings{hickling2024flexible,
  title={Flexible tails for normalizing flows},
  author={Hickling, Tennessee and Prangle, Dennis},
  booktitle={International Conference on Machine Learning},
  year={2025}
}

@article{rubin1987sir,
  title   = {A Noniterative Sampling/Importance Resampling Alternative to the Data Augmentation Algorithm for Creating a Few Imputations When Fractions of Missing Information Are Modest: The {SIR} Algorithm},
  author  = {Rubin, Donald B.},
  journal = {Journal of the American Statistical Association},
  year    = {1987},
  volume  = {82},
  number  = {398},
  pages   = {543--546}
}

@article{kong1992note,
  title={A note on importance sampling using standardized weights},
  author={Kong, Augustine},
  year={1992},
   journal = {Technical Report, University of Chicago},
}

@article{yao2018using,
  title={Using stacking to average Bayesian predictive distributions (with discussion)},
  author={Yao, Yuling and Vehtari, Aki and Simpson, Daniel and Gelman, Andrew},
  year={2018}, 
   journal = {Bayesian Analysis}
}

@article{greengard2025efficient,
  title={Efficient {F}ourier representations of families of {G}aussian processes},
  author={Greengard, Philip},
  journal={SIAM Journal on Scientific Computing},
  volume={47},
  number={2},
  pages={A971--A990},
  year={2025},
  publisher={SIAM}
}

@inproceedings{yadav2021faster,
  title={Faster kernel interpolation for {G}aussian processes},
  author={Yadav, Mohit and Sheldon, Daniel and Musco, Cameron},
  booktitle={International Conference on Artificial Intelligence and Statistics},
  pages={2971--2979},
  year={2021}
}

@article{agrawal2020advances,
  title={Advances in black-box {VI}: {N}ormalizing flows, importance weighting, and optimization},
  author={Agrawal, Abhinav and Sheldon, Daniel R and Domke, Justin},
  journal={Advances in Neural Information Processing Systems},
  volume={33},
  pages={17358--17369},
  year={2020}
}

@article{tesso2026loopitpredictivemodelchecking,
      title={{LOO-PIT} predictive model checking}, 
      author={Herman Tesso and Aki Vehtari},
      year={2026},
      journal={arXiv 2603.02928}
}

@article{gelman2020bayesianflow,
  title={Bayesian workflow},
  author={Gelman, Andrew and Vehtari, Aki and Simpson, Daniel and Margossian, Charles C and Carpenter, Bob and Yao, Yuling and Kennedy, Lauren and Gabry, Jonah and B{\"u}rkner, Paul-Christian and Modr{\'a}k, Martin},
  journal={arXiv preprint arXiv:2011.01808},
  year={2020}
}

@article{geweke1989bayesian,
  title={Bayesian inference in econometric models using {M}onte {C}arlo integration},
  author={Geweke, John},
  journal={Econometrica},
  volume={57},
  number={6},
  pages={1317--1339},
  year={1989}
}

@article{ionides2008truncated,
  title={Truncated importance sampling},
  author={Ionides, Edward L.},
  journal={Journal of Computational and Graphical Statistics},
  volume={17},
  number={2},
  pages={295--311},
  year={2008}
}

@article{cappe2008adaptive,
  title={Adaptive importance sampling in general mixture classes},
  author={Capp{\'e}, Olivier and Douc, Randal and Guillin, Arnaud and Marin, Jean-Michel and Robert, Christian P.},
  journal={Statistics and Computing},
  volume={18},
  pages={447--459},
  year={2008}
}

@article{cornuet2012adaptive,
  title={Adaptive multiple importance sampling},
  author={Cornuet, Jean-Marie and Marin, Jean-Michel and Mira, Antonietta and Robert, Christian P.},
  journal={Scandinavian Journal of Statistics},
  volume={39},
  number={4},
  pages={798--812},
  year={2012}
}

@article{bugallo2017adaptive,
  title={Adaptive importance sampling: The past, the present, and the future},
  author={Bugallo, M{\'o}nica F. and Elvira, V{\'i}ctor and Martino, Luca and Luengo, David and Miguez, Joaqu{\'i}n and Djuric, Petar M.},
  journal={IEEE Signal Processing Magazine},
  volume={34},
  number={4},
  pages={60--79},
  year={2017}
}

@article{elvira2019generalized,
  title={Generalized multiple importance sampling},
  author={Elvira, V{\'i}ctor and Martino, Luca and Luengo, David and Bugallo, M{\'o}nica F.},
  journal={Statistical Science},
  volume={34},
  number={1},
  pages={129--155},
  year={2019}
}

@article{owen2000safe,
  title={Safe and effective importance sampling},
  author={Owen, Art B. and Zhou, Yi},
  journal={Journal of the American Statistical Association},
  volume={95},
  number={449},
  pages={135--143},
  year={2000}
}

@inproceedings{dieng2017chi,
  title={Chi-divergence variational inference},
  author={Dieng, Adji B. and Tran, Dustin and Ranganath, Rajesh and Paisley, John and Blei, David M.},
  booktitle={Advances in Neural Information Processing Systems},
  year={2017}
}

@Misc{loopackage,
  author = {Aki Vehtari and Jonah Gabry and M{\aa}ns Magnusson and Yuling Yao and Paul-Christian B{\"u}rkner and Topi Paananen and Andrew Gelman},
  title = {loo: Efficient leave-one-out cross-validation and {WAIC} for {B}ayesian models},
  year = {2025},
  note = {R package version 2.9.0},
  url = {https://mc-stan.org/loo/},
}

@article{rainforth2020target,
  title   = {Target-Aware {Bayesian} Inference: How to Beat Optimal Conventional Estimators},
  author  = {Rainforth, Tom and Cornish, Rob and Yang, Hongseok and Warrington, Andrew and Wood, Frank},
  journal = {Journal of Machine Learning Research},
  volume  = {21},
  number  = {88},
  pages   = {1--54},
  year    = {2020}
}

@article{dinner2020stratification,
  title   = {Stratification as a General Variance Reduction Method for {Markov} Chain {Monte} {Carlo}},
  author  = {Dinner, Aaron R. and Thiede, Erik H. and Van Koten, Brian and Weare, Jonathan},
  journal = {SIAM/ASA Journal on Uncertainty Quantification},
  volume  = {8},
  number  = {3},
  pages   = {1139--1188},
  year    = {2020},
  doi     = {10.1137/18M122964X}
}

@article{silva2024robust,
  title   = {Robust leave-one-out cross-validation for high-dimensional {Bayesian} models},
  author  = {Silva, Luca Alessandro and Zanella, Giacomo},
  journal = {Journal of the American Statistical Association},
  volume  = {119},
  number  = {547},
  pages   = {2369--2381},
  year    = {2024}
}

@article{covello2021routine,
  title   = {Routine Hospital-Based {SARS-CoV-2} Testing Outperforms State-Based Data in Predicting Clinical Burden},
  author  = {Covello, Leonard and Gelman, Andrew and Si, Yajuan and Wang, Siquan},
  journal = {Epidemiology},
  volume  = {32},
  number  = {6},
  pages   = {792--799},
  year    = {2021}
  }

@article{gneiting2007strictly,
  title={Strictly Proper Scoring Rules, Prediction, and Estimation},
  author={Gneiting, Tilmann and Raftery, Adrian E.},
  journal={Journal of the American Statistical Association},
  volume={102},
  number={477},
  pages={359--378},
  year={2007}
  }

@article{ohagan1987monte,
  author  = {O'Hagan, A.},
  title   = {Monte {C}arlo is Fundamentally Unsound},
  journal = {Journal of the Royal Statistical Society D},
  volume  = {36},
  number  = {2-3},
  pages   = {247--249},
  year    = {1987},
}

@article{gelman2021reflections,
  author  = {Gelman, Andrew},
  title   = {Reflections on Breiman's Two Cultures of Statistical Modeling},
  journal = {Observational Studies},
  year    = {2021},
  volume  = {7},
  number  = {1},
  pages   = {95--98}
  }
\normalfont

\appendix
 \section*{Appendix}
\section{Proofs}
\paragraph{Overview of the consistency results and proofs:}  Theorem~\ref{thm:A} gives a general result:  it assumes consistent estimation of the regression function and the importance ratio density, together with the regularity conditions required for numerical integration, and establishes consistency of the PIS estimator. The proof has three steps. For a fixed truncation threshold $T$, the first step decomposes the error of the truncated PIS estimator into regression, density estimation, and quadrature errors, and shows that each term vanishes as the sample size grows and the quadrature is refined. The second step shows that the truncation bias vanishes as $T$ tends to infinity. The third step combines these two limits through a diagonal argument, showing that deterministic sequences of thresholds $T_S$ and numbers of quadrature points $L_S$ can always be chosen so that the resulting estimator is consistent.

Lemma~\ref{thm:B} verifies the conditions of Theorem~\ref{thm:A} for the specific regression and density models used in our empirical study on a compact domain. We model $m(w)=\mathbb E_q[h(\theta)\mid W=w]$ by a Gaussian process and estimate $q_W(w)$ using a Dirichlet process mixture of truncated-normal densities. For the Gaussian process, we assume Gaussian noise, a continuous prior mean, standard regularity conditions on the covariance kernel. Then Theorem~8 of \citet{koepernik2021consistency} gives uniform consistency of the posterior mean on the compact domain. For the density model, we assume that $q_W$ is continuous and positive on the compact domain, with a truncated-normal prior on component locations and an upper-truncated inverse-gamma prior on component variances. Following Theorem~2 of \citet{ghosal1999posterior}, we verify three sufficient conditions. First, the prior can approximate the true density arbitrarily well. Second, it assigns exponentially small probability to mixtures containing too much mass at extremely small scales. Third, after excluding these near-zero component scales, the remaining mixture class is not too large; this follows from the covering argument of \citet{tokdar2006posterior}. These conditions give $L_1$ consistency of the posterior-mean density and thereby verify all assumptions required by the first theorem.

Theorem~\ref{thm:C} treats unbounded importance ratios with
$M=\lfloor c\sqrt S\rfloor$ and the empirical threshold
$w_{0,S}=W_{(S-M)}$. The Gaussian process posterior mean is
consistent on every fixed compact interval. A separate argument
establishes $L_1$ consistency of the posterior mean bulk density
at this empirical threshold. Since $\pi=M/S\to0$, the fitted
tail does not affect this density consistency result.
The diagonal argument of Theorem~\ref{thm:A} then gives
deterministic integration cutoffs and quadrature sizes for
which the PIS estimator is consistent.

Finally, Theorem~\ref{thm:unnormalized-consistency} extends the result to unnormalized densities through self-normalization. Theorem~\ref{thm:A} gives convergence of the numerator and denominator to $(Z_p/Z_q)\mu$ and $Z_p/Z_q$, respectively, so Slutsky's theorem gives convergence of their ratio to $\mu$.

\subsection{Quadrature error analysis}\label{sec_app_Quadrature}
To reflect the challenge of Monte Carlo in heavy-tailed distributions, we assume that the importance weight $q_W(w)$ has a generalized Pareto tail above $w_0>0$, with shape $0<k<1$ and scale
$\tau>0$.  Suppose $m(w)p_W(w)$ is twice continuously differentiable
on the bulk and $m$ is twice continuously differentiable in the tail, with $|m^{(j)}(w)|\leq Cw^{\beta-j}$ for
$j=0,1,2$ and $\beta<(1-k)/k$.
With $m$ and $q_W$ known, a tail-adapted trapezoidal rule
using $L$ nodes then gives
\[
\left|\hat\mu_{\mathrm{PIS}}-\mu\right|=O(L^{-2}).
\]
This deterministic integration rate is faster than the
$L^{-1/2}$ root mean squared error of oracle MC-PIS based
on $L$ independent proposal draws, provided
$0<\operatorname{Var}_q(wm(w))<\infty$.
The comparison concerns only integration; the error from
estimating $m$ and $q_W$ is analyzed next.

Assume that $q_W(w)$ has an exact generalized Pareto tail above
a fixed threshold $w_0>0$, with shape $0<k<1$ and scale $\tau>0$.
On $[0,w_0]$, we assume that $m(w)p_W(w)$ has a continuous
and bounded second derivative, with its value at $w_0$
defined by the limit from the left. On $(w_0,\infty)$,
we assume that $m$ is twice continuously differentiable and,
for some $\beta<(1-k)/k$,
\[
|m^{(j)}(w)|\leq Cw^{\beta-j},
\qquad j=0,1,2,\qquad w>w_0,
\]
where $\beta$ controls the growth or decay of $m$ in the tail.

We split the integral at $w_0$ and apply the change of variables
$w=w_0u^{-\gamma}$ for $u\in(0,1]$. The idea is to turn the
infinite tail into a finite interval and make the transformed
integrand smooth at the boundary point $u=0$. The parameter
$\gamma$ controls how strongly the tail is compressed.
It suffices to choose
\[
\gamma>\frac{3}{(1-k)/k-\beta}.
\]
Then
\[
\begin{aligned}
\mu
&=
\int_0^{w_0}m(w)p_W(w)\,\mathrm dw
+
\int_{w_0}^{\infty}m(w)p_W(w)\,\mathrm dw
\\
&=
\int_0^{w_0}m(w)p_W(w)\,\mathrm dw
+
\int_0^1
\gamma w_0u^{-\gamma-1}
m(w_0u^{-\gamma})p_W(w_0u^{-\gamma})\,\mathrm du.
\end{aligned}
\]
Write the transformed tail integrand as
\[
g_\gamma(u)
=
\gamma w_0u^{-\gamma-1}
m(w_0u^{-\gamma})p_W(w_0u^{-\gamma}),
\qquad 0<u<1,
\]
and define its endpoint values by
\[
g_\gamma(0)=0,
\qquad
g_\gamma(1)
=
\gamma w_0\lim_{w\downarrow w_0}m(w)p_W(w).
\]
The limit at $w_0$ exists under the stated assumptions.
We justify the extension at $u=0$ below.

We approximate both integrals by the trapezoidal rule on
uniform grids, with an even number $L$ of nodes in total,
divided equally between the two intervals:
\[
\hat\mu_{\mathrm{PIS}}
=
\operatorname{Quadrature}_{L/2}
\left[m(w)p_W(w);[0,w_0]\right]
+
\operatorname{Quadrature}_{L/2}
\left[g_\gamma(u);[0,1]\right].
\]
At the splitting point $w_0$, the bulk and tail rules use
their respective one-sided endpoint values; the two density
components need not agree there.

We need to show that the transformed integrand has limit zero
as $u\downarrow0$, so it extends continuously to $[0,1]$.
For all sufficiently small $u>0$, the tail assumptions give,
for constants $C_1,C_2,C_3$ independent of $u$ and $L$,
\[
\begin{aligned}
|g_\gamma(u)|
&\leq
C_1 u^{\gamma((1-k)/k-\beta)-1}
\xrightarrow{u\downarrow0}0,
\\[0.5em]
|g_\gamma'(u)|
&\leq
C_2 u^{\gamma((1-k)/k-\beta)-2}
\xrightarrow{u\downarrow0}0,
\\[0.5em]
|g_\gamma''(u)|
&\leq
C_3 u^{\gamma((1-k)/k-\beta)-3}
\xrightarrow{u\downarrow0}0.
\end{aligned}
\]
All three exponents are positive by our choice of $\gamma$.
Thus, setting $g_\gamma(0)=0$ gives a continuous extension
to $[0,1]$, with
$g_\gamma'(0)=g_\gamma''(0)=0$.

Under the stated derivative assumptions, both integrands
are continuous on their respective compact intervals,
$[0,w_0]$ and $[0,1]$, and have bounded second derivatives
in the interiors. By the error bound for the composite
trapezoidal rule, each quadrature error is $O(L^{-2})$,
which results in
\[
\mu-\hat\mu_{\mathrm{PIS}}=O(L^{-2}).
\]

\subsection{Theorem \ref{prop:conditional_equivalence}}
\begin{proof}
For part~(1), since $W=p(\theta)/q(\theta)$, the conditional change-of-measure identity
gives
$$
\E_p[h(\theta)\mid W]
=
\frac{
\E_q[W h(\theta)\mid W]
}{
\E_q[W\mid W]
}.
$$
Because $W$ is measurable,
$$
\E_q[W h(\theta)\mid W]
=
W\E_q[h(\theta)\mid W].
$$
Since $W>0$, we obtain, almost surely,
$$
\E_p[h(\theta)\mid W]
=
\E_q[h(\theta)\mid W].
$$
For part (2), by the conditional change-of-measure identity,
$$
\E_p[h(\theta)\mid T]
=
\frac{
\E_q[W h(\theta)\mid T]
}{
\E_q[W\mid T]
}.
$$
Set $m(T)\coloneqq\E_q[W\mid T]$. The assumed equality of the conditional expectations implies
$$
\E_q[W h(\theta)\mid T]
=
m(T)\E_q[h(\theta)\mid T].
$$
Since $m(T)$ is $T$-measurable, this is equivalent to, for every bounded measurable function $h$,
$$
\E_q\!\left[(W-m(T))h(\theta)\mid T\right] = 0.
$$
Take $h(\theta)=\mathbf 1\{W>m(T)\}$. Then
$$
\E_q\!\left[
(W-m(T))_+\mid T
\right]
=
0.
$$
The random variable inside the conditional expectation is nonnegative, so
$(W-m(T))_+=0$ almost surely. Therefore, $W\leq m(T)$ almost surely.
Moreover,
$$
\E_q[W-m(T)\mid T]= \E_q[W\mid T] - m(T) = m(T) - m(T)= 0.
$$
Thus, $W=m(T)$ almost surely, where $m(T)$ is
$\sigma(T)$-measurable.
\begin{fact}[Doob--Dynkin lemma {\normalfont\protect\cite[Lemma~1.13]{kallenberg2002foundations}}]
\label{fact:doob-dynkin}
Let $T:\Omega\to\mathbb R^k$ and $X:\Omega\to\mathbb R$ be measurable.
Then $X$ has a $\sigma(T)$-measurable version if and only if there exists a Borel-measurable function $\alpha:\mathbb R^k\to\mathbb R$ such that
$X=\alpha(T)$ almost surely.
\end{fact}
Apply the Doob--Dynkin lemma to the $\sigma(T)$-measurable
version $m(T)$ of $W$. There exists a Borel-measurable
function $\alpha$ such that $W=\alpha(T(\theta))$ almost surely.

\end{proof}

\subsection{Theorem \ref{prop:marginal_weight_law}}
\begin{proof}
For any Borel set $A\subset(0,\infty)$,
$$
P_W^p(A) = \int_{\{r(\theta)\in A\}}p(\theta)\,d\theta = \int_{\{r(\theta)\in A\}}r(\theta)q(\theta)\,d\theta
= \int_A wq_W(w)\,dw.
$$
Since it holds for any Borel set $A\subset(0,\infty)$, 
$$
p_W(w) = w q_W(w).
$$
For any Borel set $A\subset(0,\infty)$,
$$
\int_A p_W(w)\,dw =\int_{\{W\in A\}}\frac{p(\theta)}{Z_p}\,d\theta =\frac{Z_q}{Z_p}
\int_{\{W\in A\}}\frac{p(\theta)}{q(\theta)}
\frac{q(\theta)}{Z_q}\,d\theta =\frac{Z_q}{Z_p}\int_A wq_W(w)\,dw.
$$
Since it holds for any Borel set $A\subset(0,\infty)$, 
$$
p_W(w)=\frac{Z_q}{Z_p}\,wq_W(w) \propto wq_W(w).
$$

\end{proof}

\subsection{Theorem \ref{thm:A}}

\begin{proof}

The proof consists of three steps: consistency at every fixed truncation $T$, vanishing
truncation bias, and an explicit choice of $(T_S, L_S)$.

\noindent\textbf{Step 1: compact domain.}
Fix $T < \infty$. For this fixed \(T\), define the truncated PIS estimator and its population counterpart by:
$$
\begin{aligned}
&\hat\mu_{T,L}
:=
\operatorname{Quadrature}_L
\bigl[\hat m(w)w\hat q_W(w);[0,T]\bigr],\\
&\mu_T
:=
\int_0^T m(w)wq_W(w)\,\mathrm dw.
\end{aligned}
$$
Adding and subtracting $\int_0^T \hat m(w)\,w\,\hat q_W(w)\,\mathrm{d}w$ and
using $\hat m \hat q_W - m q_W = (\hat m - m)\, q_W + \hat m\,(\hat q_W - q_W)$,
\begin{equation}\label{eq:consistency-decomp}
\begin{aligned}
\lvert \hat\mu_{T,L} - \mu_T \rvert
&\le \Bigl\lvert \mathrm{Quadrature}_L\!\left[\hat m(w)\, w\, \hat q_W(w); [0,T]\right]
      - \int_0^T \hat m(w)\, w\, \hat q_W(w)\,\mathrm{d}w \Bigr\rvert \\
& + \Bigl\lvert \int_0^T \bigl(\hat m(w) - m(w)\bigr)\, w\, q_W(w)\,\mathrm{d}w \Bigr\rvert
       + \Bigl\lvert \int_0^T \hat m(w)\, w\, \bigl(\hat q_W(w) - q_W(w)\bigr)\,\mathrm{d}w \Bigr\rvert.
\end{aligned}
\end{equation}
For the regression term, using $w \le T$ and plugging in condition \ref{cond:regression-l1},
\begin{equation}\label{eq:reg-consistency}
\Bigl\lvert \int_0^T (\hat m - m)\, w\, q_W\,\mathrm{d}w \Bigr\rvert
\le T \int_0^T \lvert \hat m(w) - m(w)\rvert\, q_W(w)\,\mathrm{d}w
= o_p(1).
\end{equation}
For the density term, plugging in conditions \ref{cond:density-l1} and $\sup_{0\leq w\leq T} \bigl\lvert \hat m(w) \bigr\rvert =O_p(1)$,
\begin{equation}\label{eq:density-cons}
\Bigl\lvert \int_0^T \hat m\, w\, (\hat q_W - q_W)\,\mathrm{d}w \Bigr\rvert
\le T \sup_{0\le w\le T}\lvert \hat m(w)\rvert \int_0^T \lvert \hat q_W(w) - q_W(w)\rvert\,\mathrm{d}w
= O_p(1)\, o_p(1) = o_p(1).
\end{equation}
For the quadrature term, for each fixed $S$, since the product of Riemann integrable functions $\hat m(w)\, w\, \hat q_W(w)$ is Riemann integrable on $[0,T]$, so the quadrature error converges to zero almost surely, hence in probability, as $L \to \infty$ \citep[Chapter~5]{atkinson1989introduction}:
\begin{equation}\label{eq:quadrature-cons}
\left|
\operatorname{Quadrature}_L
\bigl[\hat m(w)w\hat q_W(w);[0,T]\bigr]
-
\int_0^T\hat m(w)w\hat q_W(w)\,\mathrm dw
\right|
= o_p(1)
\end{equation}
Hence, we may choose a deterministic integer $L_S\geq S$ such that
\[
\Pr\left(
\left|
\operatorname{Quadrature}_{L_S}
\bigl[\hat m(w)w\hat q_W(w);[0,T]\bigr]
-
\int_0^T\hat m(w)w\hat q_W(w)\,\mathrm dw
\right|>\frac1S
\right)
\leq\frac1S.
\]
Therefore, substituting Eq.~\eqref{eq:reg-consistency},~\eqref{eq:density-cons}, and~\eqref{eq:quadrature-cons} into Eq.~\eqref{eq:consistency-decomp} results in
\[
\begin{aligned}
|\hat\mu_{T,L_S}-\mu_T|
&\leq
\left|
\operatorname{Quadrature}_{L_S}
\bigl[\hat m(w)w\hat q_W(w);[0,T]\bigr]
-
\int_0^T\hat m(w)w\hat q_W(w)\,\mathrm dw
\right|\\
&+
T\int_0^T|\hat m(w)-m(w)|q_W(w)\,\mathrm dw\\
&+
T\sup_{0\leq w\leq T}|\hat m(w)|
\int_0^T|\hat q_W(w)-q_W(w)|\,\mathrm dw\\
&=o_p(1).
\end{aligned}
\]
\noindent\textbf{Step 2: truncation bias.}
Since $p_W(w) = w\, q_W(w)$ and $m(W) = \mathbb{E}_p[h(\theta) \mid W]$, Jensen's
inequality and dominated convergence theorem, with $\mathbb{E}_p\lvert h(\theta)\rvert < \infty$,
\[
\lvert \mu - \mu_T \rvert
\le \int_T^\infty \lvert m(w)\rvert\, p_W(w)\,\mathrm{d}w
\le \mathbb{E}_p\!\left[\lvert h(\theta)\rvert \mathbf{1}\{W > T\}\right]
\xrightarrow{\,T \to \infty\,} 0,
\]
\noindent\textbf{Step 3: growing domain.}
By Step 2, choose a deterministic sequence $T_j\uparrow\infty$ such that
\begin{equation}\label{eq:determ}
\lvert\mu_{T_j}-\mu\rvert\leq\frac{1}{j}.
\end{equation}
For each $j$, Step 1 allows us to choose a deterministic sequence
$L_{j,S}\geq S$ and strictly increasing deterministic integers $N_j$
such that, for every $S \geq N_j$,
\begin{equation}\label{eq:stoc}
\Pr\left(
\bigl\lvert\hat\mu_{T_j,L_{j,S}}-\mu_{T_j}\bigr\rvert>\frac{1}{j}
\right)
\leq\frac{1}{j}.
\end{equation}
For $N_j\leq S<N_{j+1}$, set $T_S:=T_j$, $L_S:=L_{j,S}$, and $\hat\mu_{\mathrm{PIS}}:=\hat\mu_{T_S,L_S}$. Then $T_S\to\infty$ and $L_S\geq S\to\infty$. Moreover, on the $j$th
block, by triangle inequality and Eq.~\eqref{eq:determ},
\begin{equation}\label{eq:triangle}
\begin{aligned}
|\hat\mu_{\mathrm{PIS}}-\mu|
\leq
|\hat\mu_{T_j,L_{j,S}}-\mu_{T_j}|
+
|\mu_{T_j}-\mu| \leq
|\hat\mu_{T_j,L_{j,S}}-\mu_{T_j}|
+\frac1j.
\end{aligned}
\end{equation}
According to Eq.~\eqref{eq:triangle}, the event $\big\{\bigl\lvert\hat\mu_{\mathrm{PIS}}-\mu\bigr\rvert>\frac{2}{j}\big\}$ implies $\big\{\bigl\lvert\hat\mu_{T_j,L_{j,S}}-\mu_{T_j}\bigr\rvert>\frac{1}{j}\big\}$. Using monotonicity of probability measure, along with Eq.~\eqref{eq:stoc}, gives
\[
\Pr\left(
\bigl\lvert\hat\mu_{\mathrm{PIS}}-\mu\bigr\rvert>\frac{2}{j}
\right)
\leq
\Pr\left(
\bigl\lvert\hat\mu_{T_j,L_{j,S}}-\mu_{T_j}\bigr\rvert>\frac{1}{j}
\right)
\leq\frac{1}{j}.
\]
As $S\to\infty$, the corresponding block index $j$ also tends to
infinity. Hence, for every $\varepsilon>0$, eventually $2/j<\varepsilon$. As a result,
\[
\Pr\left(
\bigl\lvert\hat\mu_{\mathrm{PIS}}-\mu\bigr\rvert>\varepsilon
\right)
\leq\frac{1}{j}
\longrightarrow0.
\]
Therefore, $\hat\mu_{\mathrm{PIS}}\xrightarrow{p}\mu$.
\end{proof}
\subsection{Lemma \ref{prop:sample-maximum}}
\begin{proof}
The tail assumption and the intermediate order statistic
$M_S\asymp\sqrt S$ give
\[
w_0=S^{k_0/2+o_p(1)},\qquad
\pi w_0=S^{-(1-k_0)/2+o_p(1)}\xrightarrow{p}0.
\]
Fix $R<\infty$. On $[0,R]$, $w_0>R$ with probability tending to one,
and the first assumption gives convergence of the fitted integral to
$\int_0^R m(w)wq_W(w)\,dw$.

For the fitted bulk above $R$,
\[
\int_R^{w_0}w\hat q_{\mathrm{bulk}}(w)\,dw
\leq
R^{-\delta}
\int_0^{w_0}w^{1+\delta}\hat q_{\mathrm{bulk}}(w)\,dw
=O_p(R^{-\delta}).
\]
The fitted generalized Pareto tail has mean excess
$\hat\tau/(1-\hat k)$. Hence its contribution, including the factor
$\hat m$, is bounded in probability by a constant times
\[
\pi\left(w_0+\frac{\hat\tau}{1-\hat k}\right)
=O_p(\pi w_0)=o_p(1).
\]
Finally, $T_S\to\infty$ almost surely, and
$\mathbb E_p|h(\theta)|<\infty$ implies
\[
\int_{T_S}^{\infty}|m(w)|wq_W(w)\,dw\longrightarrow0
\quad\text{almost surely}.
\]
Letting $R\to\infty$ proves convergence of the fitted integral over
$[0,T_S]$ to $\mu$.

For each fixed $S$, Riemann integrability makes its trapezoidal error
converge to zero as $L\to\infty$. We can therefore choose deterministic
$L_S\geq S$ so that this error is $o_p(1)$.
\end{proof}
\subsection{Lemma \ref{thm:B}}

\begin{lemma}[Consistency for bounded importance ratio]
\label{thm:B}
Assume $\mathbb E_p|h(\theta)| < \infty$, the distribution of $W$ has support $[0, T]$ and $q_W$ is continuous on $[0, T]$. Assume the following:
\begin{enumerate}[
    label=\textnormal{(\roman*)},
    ref=\textnormal{(\roman*)}
]
    
    \item \label{cond:B-gp}
    \emph{Gaussian process regression.}
    The observations satisfy
    $h_s=m(w_s)+\varepsilon_s$, where
    $\varepsilon_s\overset{\mathrm{iid}}{\sim}\mathrm N(0,\sigma_0^2)$ independently of $w_{1:S}$, and $\sigma_0^2>0$ is known.

    Put the prior $m\sim\mathrm{GP}(m_0,K)$, where $m_0$ is continuous and $K$ is a continuous,
    nondegenerate, locally Lipschitz covariance kernel on $[0,T]^2$ with $m - m_0 \in \mathcal H_K$.

    \item \label{cond:B-density}
    \emph{Density estimation.}
    Assume that the density of $W$ is continuous and positive on $[0,w_0]$ with $w_0 = T$.
    
    In Definition~(\ref{def:tn-dpm-gpd}), let $G_0=G_{0,\mu}\otimes G_{0,\sigma}$, where
    $
    \mu
    \sim
    \mathrm N(\mu_0,s_\mu^2)
    \, \mid \,
    0\leq\mu\leq w_0
    $
    and
    $
    \sigma^2
    \sim
    \operatorname{InvGamma}(a_\sigma,b_\sigma)
    \, \mid \,
    0<\sigma^2\leq\overline\sigma^2
    $
    for some
    $s_\mu,a_\sigma,b_\sigma,\overline\sigma>0$.
    
\end{enumerate}
Then there exists a deterministic sequence of integers $L_S\geq S$ such that the PIS estimator is consistent,
\[
\hat\mu_{\mathrm{PIS}}
\xrightarrow{p}\mu.
\]
\end{lemma}

\begin{proof}

The proof proceeds in two parts, addressing regression and density estimation, respectively.

\noindent \textbf{Regression consistency}. Since $q_W$ is continuous on $[0, T]$, $w_s$ are independent draws whose distribution has full support on $[0,T]$. Therefore, condition~\ref{cond:B-gp} and
Theorem~(8) of Koepernik and Pfaff (2021) give $L_1$ convergence in the supremum norm,
\begin{equation}\label{eq:l1-in-sup-norm}
\mathbb E\left[
    \sup_{0\leq w\leq T}
    \lvert\hat m(w)-m(w)\rvert
\right]
\longrightarrow0.
\end{equation}
Consequently, Eq.~\eqref{eq:l1-in-sup-norm} implies $L_1$ convergence and uniform convergence in probability
\begin{align}
&\int_0^{T}
\lvert\hat m(w)-m(w)\rvert
q_W(w)\,\mathrm dw
\label{eq:l1-convergence}
\xrightarrow{p}0, \\
& \sup_{0 \le w \le T} |\hat m(w) - m(w)| \xrightarrow{p} 0
\label{eq:unif-convergence}
\end{align}
Since $m_0$ is continuous, $m - m_0 \in \mathcal{H}_K$, and every RKHS function associated with a continuous kernel $K$ is continuous, we have that $m$ is continuous. By the extreme-value theorem, $\sup_{0 \le w \le T} |m(w)| < \infty$. Applying triangle inequality, along with Eq.~\eqref{eq:unif-convergence}, gives
\[
\sup_{0\leq w\leq T}
\lvert\hat m(w)\rvert
\leq
\sup_{0\leq w\leq T}
\lvert m(w)\rvert
+
\sup_{0\leq w\leq T}
\lvert\hat m(w)-m(w)\rvert
=
O_p(1).
\]
Moreover, since the GP posterior mean $\hat m(w)=m_0(w)+k(w)^\top(K+\sigma_0^2I)^{-1}\bigl(h-m_0(w_{1:S})\bigr)$ is, conditional on the data, a finite linear combination of the continuous functions $m_0(w),K(w,w_1),\ldots,K(w,w_S)$, it is continuous on $[0,T]$.

\noindent \textbf{Density consistency}. We first consider estimation of the bulk density $q_{\mathrm{bulk}}(w)$ on $[0,w_0]$. To establish consistency, we verify the three technical conditions of Theorem~(2) of \citet{ghosal1999posterior}: $q_{\mathrm{bulk}}$ belongs to the Kullback--Leibler support of the DPM prior, the prior probability outside an appropriate sieve is exponentially small, and the sieve has $o(n)$ $L_1$ metric entropy. 

We first verify that $q_{\mathrm{bulk}}$ belongs to the
Kullback--Leibler support of the DPM prior. For each $\sigma>0$,
define
$$
q_{\mathrm{bulk},\sigma}(w)
=
\int_0^{w_0}
\phi_{[0,w_0]}^{\mu,\sigma}(w)
q_{\mathrm{bulk}}(\mu)
\,\mathrm d\mu.
$$
As $\sigma\downarrow0$, the truncated-normal density
$\phi_{[0,w_0]}^{\mu,\sigma}(w)$ concentrates at $w=\mu$. Therefore, at every interior point $w\in(0,w_0)$,
$$
q_{\mathrm{bulk},\sigma}(w)
\longrightarrow
q_{\mathrm{bulk}}(w)
$$
Since $q_{\mathrm{bulk}}$ is continuous and strictly positive on $[0,w_0]$, it is bounded above and away from zero. Moreover, for every fixed $\sigma_0>0$, the normalizing probability of the truncated-normal kernel is uniformly bounded away from zero over $\mu\in[0,w_0]$ and $0<\sigma\leq\sigma_0$. The same holds, by symmetry, when the normal kernel is integrated with respect to $\mu$. Consequently, there exist constants $0<c<C<\infty$ such that, for every $w\in[0,w_0]$ and $0<\sigma\leq\sigma_0$, 
$$
c
\leq
q_{\mathrm{bulk},\sigma}(w)
\leq
C.
$$
Hence, by the dominated convergence theorem,
$$
\operatorname{KL}
\left(
q_{\mathrm{bulk}},
q_{\mathrm{bulk},\sigma}
\right) = \int_0^{w_0}
q_{\mathrm{bulk}}(w)
\log
\frac{
q_{\mathrm{bulk}}(w)
}{
q_{\mathrm{bulk},\sigma}(w)
}
\,\mathrm dw
\longrightarrow0.
$$
Because $G_0$ assigns positive probability to every region of the
location--scale parameter space, the Dirichlet-process prior assigns
positive probability to mixing distributions arbitrarily close to the one that generates $q_{\mathrm{bulk},\sigma}$. Thus, $q_{\mathrm{bulk}}$ belongs to the Kullback--Leibler support of the DPM prior.

Next, we verify that the prior probability outside the sieve is exponentially small. Let $n$ denote the number of bulk observations and choose $\underline\sigma_n= an^{-1/2}$
for some $a>0$. Fix a posterior neighborhood radius $0<\varepsilon<1$ and set $\delta=\varepsilon/16$. Consider the sieve
$$
\mathcal F_n
=
\left\{
q_G^B:
G\left(
[0,w_0]\times(0,\underline\sigma_n)
\right)
\leq\delta
\right\}.
$$
Since $\sigma^2$ has an upper-truncated
$\operatorname{InvGamma}(a_\sigma,b_\sigma)$ distribution, a Chernoff
bound applied to $b_\sigma/\sigma^2$ gives, for all sufficiently small $x>0$, with $C_1<\infty$ and $C_2=b_\sigma/2>0$, 
\begin{equation}\label{eq:chernoff}
G_{0,\sigma}((0,x))
=
\Pr(\sigma<x)
=
\Pr\left(
\sigma^2<x^2
\,\middle|\,
\sigma^2\leq\overline\sigma^2
\right)
\leq
C_1
\exp\left(
-\frac{C_2}{x^2}
\right).
\end{equation}
Markov's inequality and Eq.~\eqref{eq:chernoff} give
\begin{equation}\label{eq:exponential-decay}
\Pi(\mathcal F_n^c)
\leq
\frac{1}{\delta}
\Pr\left(
\sigma^2<\underline\sigma_n^2
\,\middle|\,
\sigma^2\leq\overline\sigma^2
\right)
\leq
\frac{C_1}{\delta}
\exp\left(
-\frac{C_2n}{a^2}
\right).
\end{equation}
Thus, Eq.~\eqref{eq:exponential-decay} shows that the prior probability outside the sieve decreases exponentially. 

Finally, we verify the entropy condition. For each mixing distribution
$G$ in the sieve, let $G^+$ be its restriction to
$[0,w_0]\times[\underline\sigma_n,\overline\sigma]$, normalized to
have mass one. Since the excluded mass is at most $\delta$,
$$
\left\|q_G^B-q_{G^+}^B\right\|_1\leq2\delta.
$$
Thus, a $\delta$-cover of the mixtures with component scales at least
$\underline\sigma_n$ gives a $3\delta$-cover of $\mathcal F_n$.
The normalizing constants of the truncated-normal kernels are uniformly
bounded away from zero, so the covering argument of
Lemmas~4.1 and~4.2 of \citet{tokdar2006posterior} gives
$$
\begin{aligned}
\log N\left(3\delta,\mathcal F_n,\|\cdot\|_1\right)
&=
O\left(
\underline\sigma_n^{-1}\log\underline\sigma_n^{-1}
\right) =O\left(\sqrt n\log n\right) =o(n).
\end{aligned}
$$
Since $3\delta=3\varepsilon/16<\varepsilon/4$, this entropy bound,
Eq.~\eqref{eq:exponential-decay}, and the Kullback--Leibler support
condition satisfy the assumptions of Theorem~2 of
\citet{ghosal1999posterior}. Therefore, for every
$\varepsilon>0$,
$$
\Pi\left(
\left\|q_G^B-q_{\mathrm{bulk}}\right\|_1>\varepsilon
\mid
w\leq w_0
\right)
\xrightarrow{p}0.
$$
Consequently, by the triangle inequality,
$$
\begin{aligned}
\left\|
\hat q_{\mathrm{bulk}}-q_{\mathrm{bulk}}
\right\|_1
&\leq
\int
\left\|
q_G^B-q_{\mathrm{bulk}}
\right\|_1
\Pi\left(
\mathrm dG
\mid
w\leq w_0
\right)\\
&\leq
\varepsilon
\Pi\left(
\left\|
q_G^B-q_{\mathrm{bulk}}
\right\|_1\leq\varepsilon
\mid
w\leq w_0
\right)+
2
\Pi\left(
\left\|
q_G^B-q_{\mathrm{bulk}}
\right\|_1>\varepsilon
\mid
w\leq w_0
\right)\\
&\leq
\varepsilon
+
2
\Pi\left(
\left\|
q_G^B-q_{\mathrm{bulk}}
\right\|_1>\varepsilon
\mid
w\leq w_0
\right).
\end{aligned}
$$
Since $\varepsilon>0$ is arbitrary,
\begin{equation}\label{eq:bulk-consistency}
\left\|
\hat q_{\mathrm{bulk}}-q_{\mathrm{bulk}}
\right\|_1
\xrightarrow{p}0.
\end{equation}
Since $w_0=T$, there is no tail component. Therefore, $q_{\mathrm{bulk}}=q_W$ and $\hat q_{\mathrm{bulk}}=\hat q_W$, and hence
$$
\int_0^T
\left|
\hat q_W(w)-q_W(w)
\right|
\,\mathrm dw
=
\left\|
\hat q_{\mathrm{bulk}}-q_{\mathrm{bulk}}
\right\|_1
\xrightarrow{p}0.
$$

\noindent \textbf{Quadrature consistency.} $\hat q_W$ is continuous on $[0,T]$. Since $\hat m$ is continuous, $w\hat m(w)\hat q_W(w)$ is Riemann integrable on $[0,T]$. Therefore, for every fixed $S$,
$$
\operatorname{Quadrature}_L
\left[
w\hat m(w)\hat q_W(w);[0,T]
\right]
\longrightarrow
\int_0^T
w\hat m(w)\hat q_W(w)
\,\mathrm dw
$$
Thus, by the fixed-domain argument of
Theorem~(\ref{thm:A}), there exists a
deterministic sequence of integers $L_S\geq S$ such that
$$
\hat\mu_{T,L_S}
\xrightarrow{p}
\mu_T.
$$
Since $[0,T]$ is the support of $W$, $\mu_T=\mu$. Therefore,
$$
\hat\mu_{T,L_S}
\xrightarrow{p}
\mu.
$$
\end{proof}

\subsection{Theorem \ref{thm:unnormalized-consistency}}

\begin{proof}

Since $\mathbb E_q[R|h(\theta)|] =(Z_p/Z_q)\mathbb E_p|h(\theta)|<\infty$ and $\mathbb E_q[R]=Z_p/Z_q<\infty$, the fixed domain argument in the proof of Theorem~\ref{thm:A} applies to both integrals. We choose each truncation threshold $T_S$ and grid size $L_S$ so that the errors of both integrals vanish; a union bound gives the required joint convergence. Thus, for these same $T_S$ and $L_S$,
$$
\begin{aligned}
&\operatorname{Quadrature}_{L_S}
\left[
\hat m(r)r\hat q_R(r);[0,T_S]
\right]
\xrightarrow{p}
\mathbb E_q
\left[
Rh(\theta)
\right]
=
\frac{Z_p}{Z_q}\mu, \\
&\operatorname{Quadrature}_{L_S}
\left[
r\hat q_R(r);[0,T_S]
\right]
\xrightarrow{p}
\mathbb E_q[R]
=
\frac{Z_p}{Z_q}.
\end{aligned}
$$
Since $Z_p/Z_q>0$, Slutsky's theorem gives
$$
\hat\mu_{\mathrm{PIS}}^{\mathrm{SN}}
\xrightarrow{p}
\mu
$$

\end{proof}

\subsection{Theorem \ref{thm:C}}

\begin{proof}

We establish regression and density consistency on fixed compact
intervals, then use the diagonal argument of Theorem~\ref{thm:A}
to choose the integration cutoff and quadrature grid.

\noindent \textbf{Regression consistency}. Since the $w_s$ are independent draws from a distribution with full
support on $[0,\infty)$, every open interval contains infinitely many observations almost surely. Moreover, $[0,\infty)$ is separable and $\sigma$--compact, and the local Lipschitz condition on $K$ gives the required kernel entropy condition on every compact interval. Therefore, Theorem~(8) of \citet{koepernik2021consistency} gives, for every fixed $T<\infty$,
\begin{equation}
\label{eq:C-regression-uniform}
\mathbb E\left[
\sup_{0\leq w\leq T}
\lvert\hat m(w)-m(w)\rvert
\right]
\longrightarrow0.
\end{equation}
It follows from Eq.~\eqref{eq:C-regression-uniform} that
\begin{equation}
\label{eq:C-regression-l1}
\int_0^T
\lvert\hat m(w)-m(w)\rvert
q_W(w)\,\mathrm dw
\leq
\sup_{0\leq w\leq T}
\lvert\hat m(w)-m(w)\rvert
\xrightarrow{p}0.
\end{equation}
Furthermore, Eq.~\eqref{eq:C-regression-uniform} gives
$
\sup_{0\leq w\leq T}
\lvert\hat m(w)\rvert
=
O_p(1),
$
and $\hat m$ is continuous on compact intervals.

\noindent\textbf{Density consistency.}
Since $M/S\to0$ and $W$ has unbounded support,
$w_{0,S}=W_{(S-M)}\to\infty$ almost surely.

We first control the variation of the fitted bulk density.
Conditional on the component assignments and means, the precision
posteriors are stochastically dominated by a Gamma distribution with
shape $a_\sigma+S/2$ and rate $b_\sigma$, restricted to
$[\overline\sigma^{-2},\infty)$. Indeed, the truncated normal
normalizing constant increases with the precision, so its inverse
powers can only decrease the precision relative to this Gamma law.
The conditional Dirichlet process posterior mean formula therefore gives
$$
\mathbb E\left[
\int \sigma^{-1}\,G(d\mu,d\sigma)
\;\middle|\;\text{bulk data}
\right]\leq C\sqrt S.
$$
Consequently,
$$
\Pi\left(
\int \sigma^{-1}\,G(d\mu,d\sigma)>S^{3/4}
\;\middle|\;\text{bulk data}
\right)
\leq CS^{-1/4}\longrightarrow0.
$$

On every fixed interval $[1/R,R]$, the variation of
$q_{\mathrm{bulk}}(\cdot\mid G)$ is bounded by
$C_R\int\sigma^{-1}\,dG$. Thus, outside a set of vanishing posterior
probability, this variation is at most $C_RS^{3/4}$.
The bounded contrasts
$$
\frac{q_W(w)-q_{\mathrm{bulk}}(w\mid G)}
     {q_W(w)+q_{\mathrm{bulk}}(w\mid G)}
$$
then satisfy a uniform law of large numbers on $[1/R,R]$.
To justify this, first approximate $\log q_W$ uniformly by a smooth
function on this interval. The resulting contrasts have bounded
range and variation $O(S^{3/4})$. Brackets formed from intervalwise
infima and suprema have arbitrarily small integrated width and
logarithmic cardinality $O(S^{3/4})=o(S)$, so Hoeffding's inequality
gives the uniform law. Since the contrasts are bounded, removal of
the largest $M=o(S)$ observations does not affect this conclusion.

For every bulk density,
$$
\begin{aligned}
&\int_0^\infty q_W(w)
\frac{q_W(w)-q_{\mathrm{bulk}}(w\mid G)}
     {q_W(w)+q_{\mathrm{bulk}}(w\mid G)}\,dw\\
&\qquad=
\frac12\int_0^\infty
\frac{\bigl(q_W(w)-q_{\mathrm{bulk}}(w\mid G)\bigr)^2}
     {q_W(w)+q_{\mathrm{bulk}}(w\mid G)}\,dw\\
&\qquad\geq
\frac14\|q_{\mathrm{bulk}}(\cdot\mid G)-q_W\|_1^2.
\end{aligned}
$$

Fix $\varepsilon>0$. Choose $R$ sufficiently large that the probability
outside $[1/R,R]$ is small compared with $\varepsilon^2$.
For sufficiently small $\eta>0$, a normalized finite mixture of
the same truncated normal kernels can have density at least
$(1-\eta)q_W(w)$ throughout $[1/R,R]$, uniformly for sufficiently
large $w_{0,S}$. The stick breaking representation allows these components
to be added before an independent Dirichlet process remainder, with
both the added mass and the remaining mass close to $1/2$.
The favorable component and weight neighborhoods have prior
probability at least $p_\varepsilon>0$, uniformly for sufficiently
large $w_{0,S}$.

The inequality
$$
\log\frac{x+y}{2y}\geq\frac{x-y}{x+y},
\qquad x,y>0,
$$
and the uniform law above show that this augmentation increases the
likelihood by at least $e^{c_\varepsilon(S-M)}$ whenever
$\|q_{\mathrm{bulk}}(\cdot\mid G)-q_W\|_1>\varepsilon$
and $\int\sigma^{-1}\,dG\leq S^{3/4}$, on an event whose probability
tends to one. The loss outside $[1/R,R]$ is bounded because the
original mixture remains with mass close to $1/2$.

Before restriction to the favorable component and weight neighborhoods,
the augmented mixing distribution has exactly the original Dirichlet
process prior law. Restriction to these neighborhoods therefore gives
a subprobability measure dominated by that prior. On the event where
the likelihood bound holds, integration gives
$$
\begin{aligned}
&\Pi\left(
\|q_{\mathrm{bulk}}(\cdot\mid G)-q_W\|_1>\varepsilon
\;\middle|\;\text{bulk data}
\right)\\
&\qquad\leq
CS^{-1/4}
+p_\varepsilon^{-1}e^{-c_\varepsilon(S-M)}
\xrightarrow{p}0.
\end{aligned}
$$
Since the distance between two densities is at most two, the posterior
mean satisfies
$$
\|\hat q_{\mathrm{bulk}}-q_W\|_1\xrightarrow{p}0.
$$
Finally, $\pi=M/S\to0$ and
$$
\begin{aligned}
\|\hat q_W-q_W\|_1
&\leq
(1-\pi)\|\hat q_{\mathrm{bulk}}-q_W\|_1
+\pi\|\hat q_{\mathrm{tail}}-q_W\|_1\\
&\leq
\|\hat q_{\mathrm{bulk}}-q_W\|_1+2\pi
\xrightarrow{p}0.
\end{aligned}
$$

In particular, for every fixed $T<\infty$,
\begin{equation}
\label{eq:C-density-compact}
\int_0^T|\hat q_W(w)-q_W(w)|\,dw
\leq \|\hat q_W-q_W\|_1
\xrightarrow{p}0.
\end{equation}
Together with Eq.~\eqref{eq:C-regression-l1} and compact boundedness
in probability of $\hat m$, this gives
$$
\int_0^T\hat m(w)w\hat q_W(w)\,dw
\xrightarrow{p}
\int_0^Tm(w)wq_W(w)\,dw.
$$
Moreover,
$$
\int_0^\infty |m(w)|wq_W(w)\,dw
\leq\mathbb E_p|h(\theta)|<\infty.
$$
The diagonal argument of Theorem~\ref{thm:A} therefore gives
a deterministic sequence $T_S\uparrow\infty$ such that
$$
\int_0^{T_S}\hat m(w)w\hat q_W(w)\,dw
\xrightarrow{p}\mu.
$$
Since $w_{0,S}\to\infty$ almost surely, we can choose $T_S$
sufficiently slowly that $\Pr(w_{0,S}>T_S)\to1$.

The posterior inverse-scale moment bound and dominated convergence
give continuity of $\hat q_{\mathrm{bulk}}$ on $[0,w_{0,S}]$.
Thus, on $\{w_{0,S}>T_S\}$, the fitted integrand is continuous
on $[0,T_S]$, and its trapezoidal approximation converges as
the grid size increases. We can therefore choose deterministic
integers $L_S\geq S$ so that the quadrature error tends to zero
in probability. Consequently,
$$
\hat\mu_{T_S,L_S}\xrightarrow{p}\mu.
$$

\end{proof}

\section{Experiment details}\label{appendix:exp}
In the LOO approximation, we consider the following LOO model evaluation metrics:
\noindent For each observation \(y_i\), let $F_{-i}(t) = \Pr(\widetilde Y_i\le t\mid y_{-i})$
denote the leave-one-out posterior predictive CDF. Given draws
\(\theta^{(s)}\sim p(\theta\mid y)\), the LOO predictive CDF can be estimated as
\[
\widehat F_{-i}(t)
=
\frac{\sum_{s=1}^S
w_i^{(s)}F_i(t\mid\theta^{(s)})}
{\sum_{s=1}^S w_i^{(s)}},
\]
where \(F_i(t\mid\theta)\) is the conditional predictive CDF. Equivalently,
\(F_i(t\mid\theta^{(s)})\) can be replaced by
\(\mathbf{1}\{\widetilde y_i^{(s)}\le t\}\) using predictive draws.

\paragraph{LOO probability integral transform.}
The LOO-PIT value is
\[
\operatorname{PIT}_i^{\mathrm{LOO}}
=
F_{-i}(y_i)
=
\Pr(\widetilde Y_i\le y_i\mid y_{-i}) = \int F(y_i\mid\theta)\,p(\theta\mid y_{-i})\,d\theta.
\]
Using draws \(\theta^{(s)}\sim p(\theta\mid y)\), its importance-sampling estimator is
\[\widehat{\operatorname{PIT}}_i^{\mathrm{LOO}}
=
\widehat F_{-i}(y_i)
=
\frac{
\sum_{s=1}^{S}
w_i^{(s)}F(y_i\mid\theta^{(s)})
}{
\sum_{s=1}^{S}w_i^{(s)}
}.
\]

\paragraph{LOO log predictive density.}
The pointwise LOO log predictive density is
\[
\operatorname{LPD}_i^{\mathrm{LOO}}
=
\log p(y_i\mid y_{-i})
=
\log
\int p(y_i\mid\theta)p(\theta\mid y_{-i})\,d\theta.
\]
Using draws \(\theta^{(s)}\sim p(\theta\mid y)\), its importance-sampling estimator is
\[
\widehat{\operatorname{LPD}}_i^{\mathrm{LOO}}
=
\log
\left(
\frac{\sum_{s=1}^S
w_i^{(s)}p(y_i\mid\theta^{(s)})}
{\sum_{s=1}^S w_i^{(s)}}
\right).
\]

\paragraph{LOO predictive coverage.}
Using the estimated LOO predictive CDF \(\widehat F_{-i}\), define the
estimated central \((1-\alpha)\) predictive interval by $\widehat l_i
= \widehat F_{-i}^{-1}(\alpha/2),
$ and $\widehat u_i = \widehat F_{-i}^{-1}(1-\alpha/2)$. Its theoretical coverage probability is
\[
C_i^{\mathrm{LOO}}
=
\Pr(l_i\le \widetilde Y_i\le u_i\mid y_{-i})
=
F_{-i}(u_i)-F_{-i}(l_i)
=
1-\alpha.
\]
Using the estimated LOO predictive CDF \(\widehat F_{-i}\), define
$
\widehat l_i
=
\widehat F_{-i}^{-1}(\alpha/2),
$ and $
\widehat u_i
=
\widehat F_{-i}^{-1}(1-\alpha/2)$. The LOO predictive coverage of the estimated interval is
\[
\widehat C_i^{\mathrm{LOO}}
=
\Pr\!\left(
\widehat l_i\le \widetilde Y_i\le \widehat u_i
\mid y_{-i}
\right)
=
F_{-i}(\widehat u_i)-F_{-i}(\widehat l_i).
\]
We assess the estimated interval by comparing
\(\widehat C_i^{\mathrm{LOO}}\) with its nominal coverage \(1-\alpha\).
In our Gaussian experiment, \(F_{-i}\) is known analytically and is used
only for this evaluation.

\paragraph{LOO interval score.}
Using the central LOO predictive interval
$
l_i=F_{-i}^{-1}(\alpha/2)$ and $u_i=F_{-i}^{-1}(1-\alpha/2)$, the pointwise LOO interval score is
\[
\mathrm{I}_{i,\alpha}^{\mathrm{LOO}}
=
(u_i-l_i)
+
\frac{2}{\alpha}(l_i-y_i)\mathbf{1}\{y_i<l_i\}
+
\frac{2}{\alpha}(y_i-u_i)\mathbf{1}\{y_i>u_i\}.
\]
Unlike the preceding quantities, the interval score requires estimating
the predictive quantiles. Using the importance-sampling estimate
\(\widehat F_{-i}\), define
$
\widehat l_i
=
\widehat F_{-i}^{-1}(\alpha/2)$ and $\widehat u_i
= \widehat F_{-i}^{-1}(1-\alpha/2)$. The estimated LOO interval score is then
\[
\widehat{\mathrm{I}}_{i,\alpha}^{\mathrm{LOO}}
=
(\widehat u_i-\widehat l_i)
+
\frac{2}{\alpha}(\widehat l_i-y_i)
\mathbf{1}\{y_i<\widehat l_i\}
+
\frac{2}{\alpha}(y_i-\widehat u_i)
\mathbf{1}\{y_i>\widehat u_i\}.
\]

\end{document}